\documentclass[11pt]{amsart}

\usepackage{amssymb}
\usepackage{amsfonts}
\usepackage{indentfirst}
\usepackage{amsopn}
\usepackage{amsmath}
\usepackage{amsthm}
\usepackage{amscd}
\usepackage{graphicx}
\usepackage{epstopdf}

\def\theequation {\thesection.\arabic{equation}}
\makeatletter\@addtoreset {equation}{section}\makeatother

\begin{document}

\title[Rogue waves on the double-periodic background]{\bf Rogue waves on the double-periodic background \\
in the focusing nonlinear Schr\"{o}dinger equation}

\author{Jinbing Chen}
\address{School of Mathematics, Southeast University, Nanjing,
Jiangsu 210096, P.R. China}

\author{Dmitry E. Pelinovsky}
\address{Department of Mathematics, McMaster University,
Hamilton, Ontario, Canada, L8S 4K1}

\author{Robert E. White}
\address{Department of Mathematics, McMaster University, Hamilton,
Ontario, Canada, L8S 4K1}

\date{\today}
\begin{abstract} 
	The double-periodic solutions of the focusing nonlinear Schr\"{o}dinger
	equation
	have been previously obtained by the method of separation of variables. We
	construct these solutions
	by using an algebraic method with two eigenvalues. Furthermore, we
	characterize the Lax spectrum for
	the double-periodic solutions and analyze rogue waves arising on their
	background. Magnification of the rogue waves is
	studied numerically.
\end{abstract}
\maketitle

\section{Introduction}

Rogue waves are commonly defined as gigantic waves appearing from nowhere and
disappearing
without trace. They are frequently seen on the ocean's surface \cite{Charif}
and in optical fibers \cite{Wabnitz}.  Their appearance is related to the
modulation instability of the wave background \cite{ZakOst}. Formation of particular rogue waves
such as Akhmediev breathers, Peregrine solution, and Kuznetsov--Ma breathers have been
modeled from different initial data such as local condensates \cite{Gelash1},
multi-soliton gases \cite{Gelash2,ZakGelash}, and periodic perturbations \cite{AZ1,AZ2}.
Experimental observations of rogue waves have been confirmed both in hydrodynamical
and optical laboratories \cite{experiments1,breathturb}.
Statistical analysis of rogue waves has recently been developed \cite{experiments2} (see also \cite{AZ1,Gelash2}).

Rogue waves and modulation instability of the wave background are commonly modeled
by the focusing nonlinear Schr\"{o}dinger (NLS) equation,
which we take in the following dimensionless form:
\begin{equation}
i \psi_t + \frac{1}{2} \psi_{xx} + |\psi|^2 \psi = 0.
\label{nls}
\end{equation}
This fundamental model is rich of many exact solutions due to its integrability discovered in \cite{ZS}.

A number of important new results were recently obtained in the mathematical theory of rogue waves.
Universal behavior of the modulationally unstable constant background was studied asymptotically in
\cite{Biondini}. The finite-gap method was employed to relate the unstable modes on the constant background
with the occurrence of rogue waves \cite{GS1,GS2}. Rogue waves of infinite
order were constructed in \cite{Bil3} based on recent developments in the inverse scattering method \cite{Bil2}.
Rogue waves of the soliton superposition were studied asymptotically in the limit of many solitons \cite{Bil1,PelSl}.

Many wave patterns are periodic in space and time variables. Simplest
travelling wave solutions have a space-periodic and time-independent profile of $|\psi|$.
Modulational instability of travelling wave solutions was analyzed in many explicit details \cite{DS,DU}.
Numerical experiments showed formation of rogue waves from modulationally unstable travelling waves \cite{AZ2}.
Exact solutions for rogue waves on the travelling wave background were constructed in our previous work  \cite{CPnls,CPWnls}
by using an algebraic method with one eigenvalue. Similar exact solutions for
rogue waves were
constructed in \cite{Feng}.

\begin{figure}[h!]
	\centering
	\includegraphics[width=8.5cm,height=6.5cm]{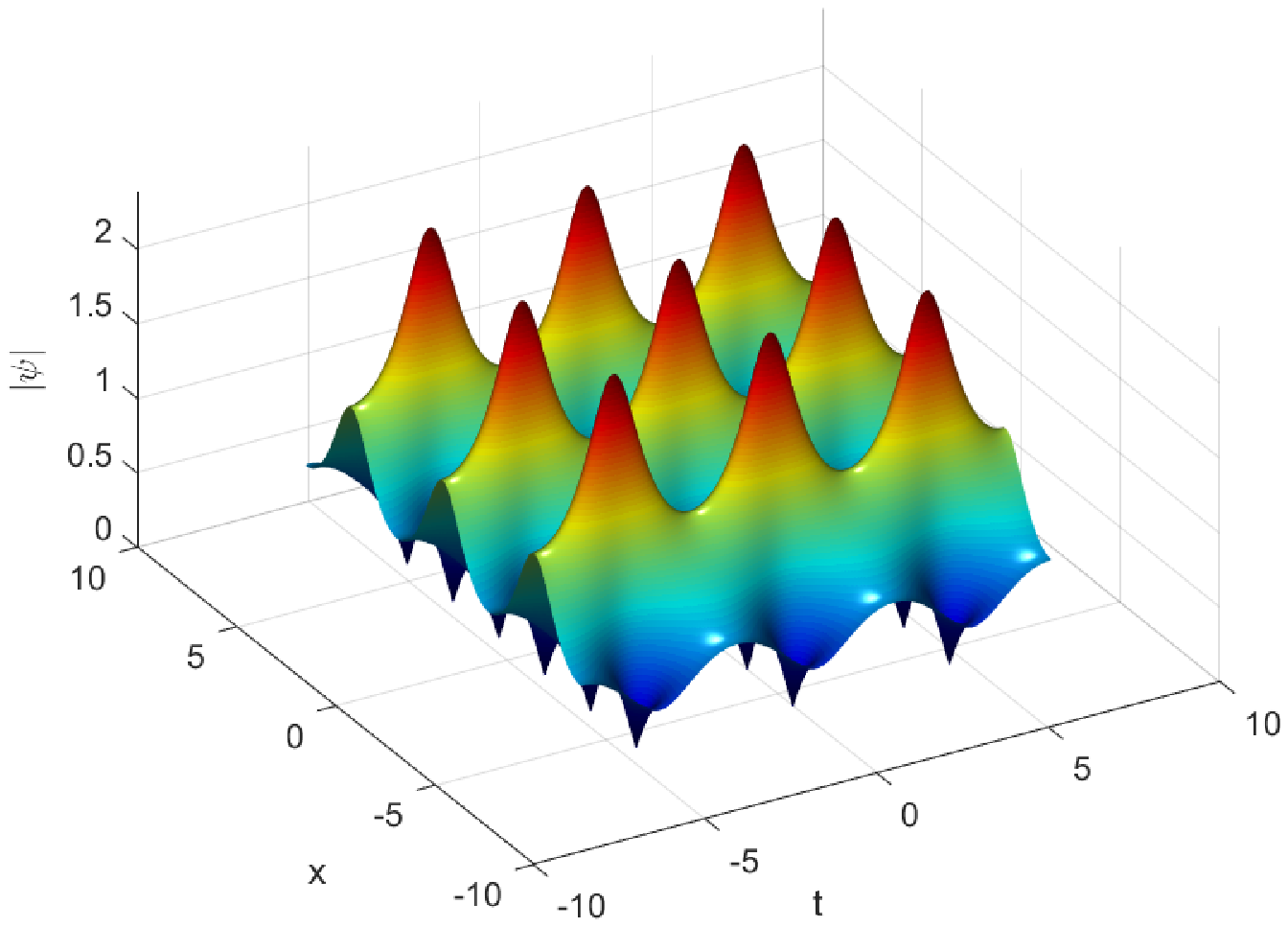}
	\includegraphics[width=8.5cm,height=6.5cm]{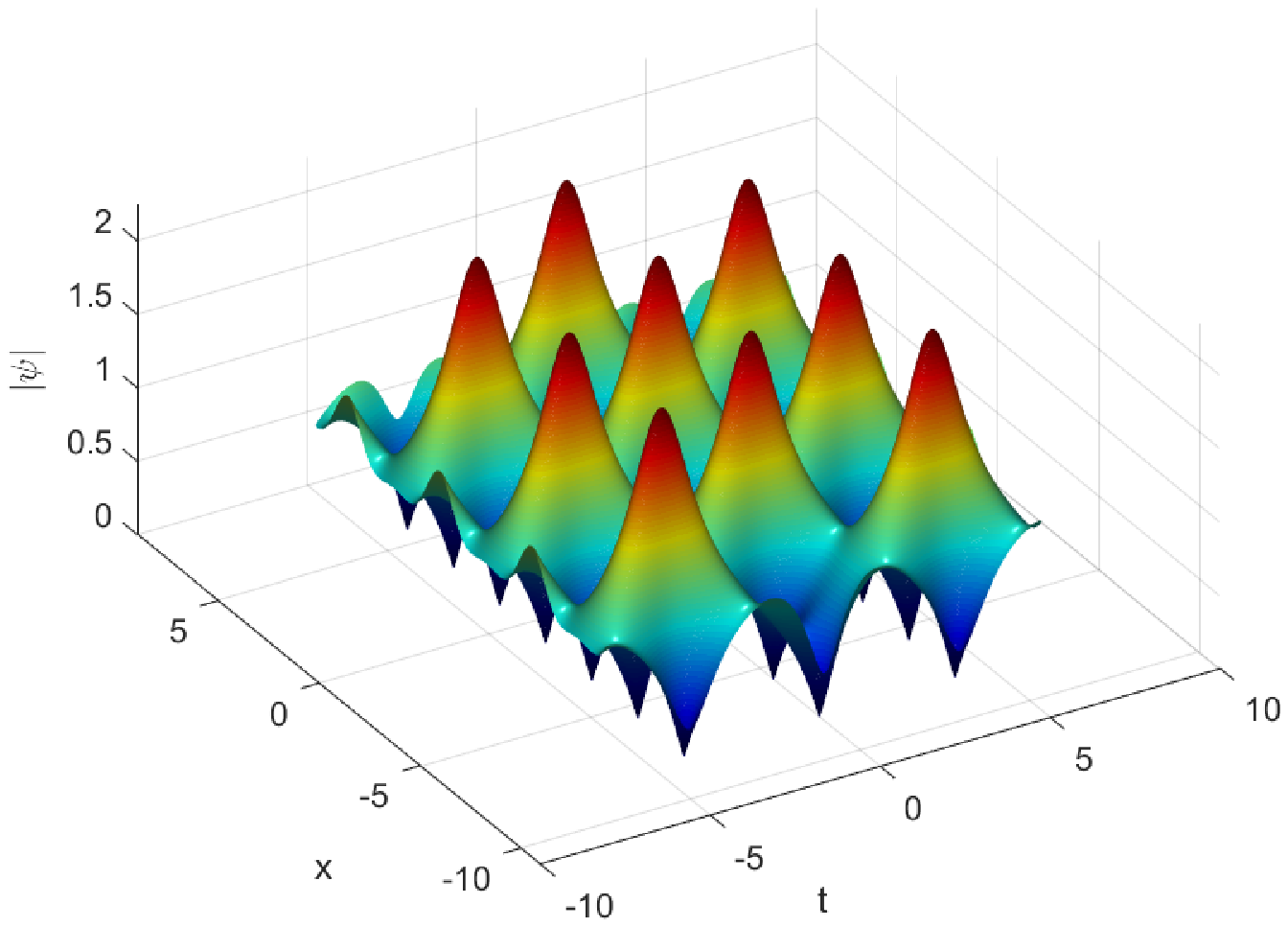}
	\caption{Solution surface for the double-periodic solution (\ref{solB}) with $k=0.9$ (left) and
        the double-periodic solution (\ref{solA}) with $k=0.8$ (right).}
	\label{f1}
\end{figure}

There exist other exact solutions of the focusing NLS equation (\ref{nls}) for which $|\psi|$ is periodic both in space and in time.
These solutions describe spatial-temporal wave patterns and are referred to as {\em the double-periodic background}.
Exact solutions for the double-periodic background were constructed in \cite{Nail1} by separating the variables and
reducing the NLS equation (\ref{nls}) to the first-order quadratures. Two
families of such solutions are given by the following rational functions of
Jacobian elliptic functions ${\rm sn}$, ${\rm cn}$, and ${\rm dn}$:
\begin{equation}
\label{solB}
\psi(x,t) = k \frac{{\rm cn}(t;k) {\rm cn}(\sqrt{1+k}x;\kappa) + i \sqrt{1+k}
{\rm sn}(t;k) {\rm dn}(\sqrt{1+k} x; \kappa)}{\sqrt{1+k} {\rm
dn}(\sqrt{1+k}x;\kappa) - {\rm dn}(t;k) {\rm cn}(\sqrt{1+k} x; \kappa)} e^{i
t}, \quad \kappa = \frac{\sqrt{1-k}}{\sqrt{1+k}}
\end{equation}
and
\begin{equation}
\label{solA}
\psi(x,t) = \frac{{\rm dn}(t;k) {\rm cn}(\sqrt{2} x; \kappa) + i \sqrt{k(1+k)} {\rm sn}(t;k)}{\sqrt{1+k}
- \sqrt{k} {\rm cn}(t;k) {\rm cn}(\sqrt{2} x; \kappa)} e^{i k t}, \quad \kappa
= \frac{\sqrt{1-k}}{\sqrt{2}},
\end{equation}
where $k \in (0,1)$.  It follows from (\ref{solB}) and (\ref{solA}) that
\begin{equation}
\label{periodicity}
|\psi(x,t)| = |\psi(x+L,t)| = |\psi(x,t+T)|, \quad (x,t) \in \mathbb{R}^2
\end{equation}
with the fundamental periods $L = \frac{4 K(\kappa)}{\sqrt{1+k}}$ and $T = 2K(k)$ for (\ref{solB})
and $L = 2 \sqrt{2} K(\kappa)$ and $T = 4 K(k)$ for (\ref{solA}),
where $K(k)$ denotes the complete elliptic integral of the first kind with
the elliptic modulus parameter $k$. Figure \ref{f1} shows the double-periodic solution \eqref{solB} for $k = 0.9$
(left) and the double-periodic solution \eqref{solA} for $k = 0.8$ (right).

The exact solutions (\ref{solB}) and (\ref{solA}) were used to describe
transformations of continuous waves
into trains of pulses \cite{Nail2}. These solutions were also constructed for the Hirota equation and
other higher-order NLS equations \cite{Nail3}. Rogue waves on the background of the double-periodic solution (\ref{solB}) were constructed
numerically in \cite{CalSch} by using numerical approximations of eigenfunctions of the Lax spectrum and the one-fold Darboux transformation.

The main purpose of this work is to construct rogue waves on the double-periodic background in a closed analytical form
and to study their magnification factors. We use a general method of nonlinearization of Lax equations
(originally proposed in \cite{Cao1} and developed for the NLS
equation in \cite{ZhouJMP,ZhouStudies}) and apply it to the case of two eigenvalues compared to the case of
one eigenvalue considered in \cite{CPnls,CPWnls}.
Although the exact solutions for rogue waves in the focusing NLS equation are
more complicated than
the corresponding solutions for the modified Korteweg-de Vries equation (which have been analyzed in \cite{CPkdv,CPgardner}),
efficient computational methods are developed to visualize the Lax spectrum associated with the double-periodic
solutions, the admissible eigenvalues, and rogue waves on the double-periodic background.
We also study magnification of such rogue waves that depends on parameters of the double-periodic background.

The conventional definition of the rogue wave's magnification factor is {\em the ratio of the maximal value
of the wave amplitude to the average value of the wave background}. The wave is considered to be rogue
if the ratio exceeds the double factor \cite{Tovbis1}. In a series of mathematical papers, the magnification factors
were analyzed for general quasi-periodic solutions of the focusing NLS equation (\ref{nls}) given
by the Riemann Theta functions in the case of genus two \cite{Wright}
and of arbitrary genus \cite{Tovbis2,Wright2}. The double-periodic solutions (\ref{solB}) and (\ref{solA})
appear to be particular cases of the Riemann Theta functions of genus two, however, the rogue waves constructed
in our paper correspond to degenerate cases of the Riemann Theta functions of genus three.
We show that the magnification factors for these rogue waves exceed the triple
factor and that these rogue waves represent isolated waves on the
double-periodic background on the $(x,t)$ plane .
We also show that if the magnification factor is defined as {\em the ratio of the maximal value of the wave amplitude
to the maximal value of the wave background}, then the magnification factor does not exceed the triple factor
achieved by the Peregrine solution on the constant background.

The paper is organized as follows. An algebraic method with two eigenvalues is developed in Section \ref{section-2},
where we derive a general fourth-order Lax--Novikov equation for the NLS equation (\ref{nls}).
In Section \ref{section-3}, we recover the double-periodic solutions (\ref{solB}) and (\ref{solA})
from reduction of the fourth-order Lax--Novikov equation to the third-order Lax--Novikov equation and its integration in first-order quadratures.
We also compute eigenvalues of the Lax spectrum analytically and
the spectral bands of the Lax spectrum numerically. Rogue waves on the double-periodic background are
constructed in Section \ref{sec-eigenvectors}. We show numerically that the rogue waves are isolated in the $(x,t)$ plane
and plot the solution surfaces for the rogue waves. The magnification factor is investigated with respect to
parameters of the rogue wave solutions and we show numerically that it exceeds the triple factor.
Section \ref{sec-conclusion} contains conclusions and further directions of study.
Appendices A and B give technical details on how the analytical expressions (\ref{solB}) and (\ref{solA})
are deduced from the general solution of the third-order Lax--Novikov equation.

\section{Algebraic method with two eigenvalues}
\label{section-2}

A solution $\psi = u$ to the NLS equation (\ref{nls}) is a compatibility
condition ($\varphi_{xt} = \varphi_{tx}$)
of the following pair of linear equations on $\varphi \in \mathbb{C}^2$:
\begin{equation}\label{3.1}
\varphi_x = U(\lambda,u) \varphi,\qquad \qquad
U(\lambda,u) = \left(\begin{array}{cc} \lambda & u \\ -\bar{u} & -\lambda
\end{array} \right)
\end{equation}
and
\begin{equation}\label{3.2}
\varphi_t = V(\lambda,u) \varphi,\qquad
V(\lambda,u) = i \left(\begin{array}{cc}
\lambda^2 + \frac{1}{2} |u|^2 & \frac{1}{2} u_x + \lambda u\\
\frac{1}{2} \bar{u}_x - \lambda\bar{u} & -\lambda^2 - \frac{1}{2} |u|^2\\
\end{array}
\right),
\end{equation}
where $\bar{u}$ is the conjugate of $u$ and $\lambda \in \mathbb{C}$ is a spectral parameter.

The procedure of computing a new solution $\psi = \hat{u}$
to the NLS equation (\ref{nls}) from another solution $\psi = u$ is well-known \cite{CalSch,CPnls,CPWnls,Feng}.
Let $\varphi = (p_1,q_1)^t$ be any nonzero solution to the linear equations (\ref{3.1}) and (\ref{3.2})
for a fixed value $\lambda = \lambda_1$. The new solution is given by the one-fold Darboux transformation
\begin{equation}
\label{1-fold}
\hat{u} = u + \frac{2 (\lambda_1 + \bar{\lambda}_1) p_1 \bar{q}_1}{|p_1|^2 +
|q_1|^2}.
\end{equation}
If $u$ is a double-periodic solution, then $\hat{u}$ may represent a rogue wave on the double-periodic background.
The main question is which value $\lambda_1$ to fix and which nonzero solution $\varphi$ to the linear equations
(\ref{3.1})--(\ref{3.2}) to take. We show that the admissible values of $\lambda_1$ for the double-periodic solutions
(\ref{solB}) and (\ref{solA}) are defined by the algebraic method with two eigenvalues.
The latter is a particular case of a more general method of nonlinearization
of the linear equations on $\varphi$, see \cite{ZhouJMP,ZhouStudies}.

\subsection{Nonlinearization of the linear equations}

Let $\varphi = (p_1,q_1)^t$ and $\varphi = (p_2,q_2)^t$ be two nonzero solutions
of the linear equations (\ref{3.1})--(\ref{3.2}) with fixed $\lambda = \lambda_1$ and
$\lambda = \lambda_2$ such that $\lambda_1 + \bar{\lambda}_1 \neq 0$, $\lambda_2 + \bar{\lambda}_2 \neq 0$,
$\lambda_1 + \bar{\lambda}_2 \neq 0$, and $\lambda_1 \neq \lambda_2$.
The following notations will be useful in this work:
\begin{eqnarray*}
\textsf{p} & := & (p_1,p_2,\bar{q}_1,\bar{q}_2)^t,\\
\textsf{q} & := & (q_1,q_2,-\bar{p}_1,-\bar{p}_2)^t,
\end{eqnarray*}
and
$$
\Lambda := {\rm diag}(\lambda_1,\lambda_2,-\bar{\lambda}_1,-\bar{\lambda}_2).
$$
Following \cite{ZhouJMP,ZhouStudies}, we introduce the following relation between
the solution $u$ to the NLS equation (\ref{nls}) and the squared eigenfunctions:
\begin{equation}\label{3.3}
u = \langle \textsf{p}, \textsf{p} \rangle :=
p_1^2+p_2^2+\bar{q}_1^2+\bar{q}_2^2.
\end{equation}
The spectral problem (\ref{3.1}) is nonlinearized into the Hamiltonian system given by
\begin{equation}\label{3.4}
\frac{d p_j}{d x} = \frac{\partial H}{\partial q_j}, \quad
\frac{d q_j}{d x} = - \frac{\partial H}{\partial p_j}, \quad j = 1,2,
\end{equation}
where
\begin{equation}\label{3.5}
H =\langle\Lambda \textsf{p}, \textsf{q} \rangle+\frac12\langle \textsf{p}, \textsf{p} \rangle
\langle \textsf{q},\textsf{q}\rangle.
\end{equation}
The time-evolution problem (\ref{3.2}) is nonlinearized into another Hamiltonian system given by
\begin{equation}\label{3.4a}
\frac{d p_j}{d t} = \frac{\partial K}{\partial q_j}, \quad
\frac{d q_j}{d t} = - \frac{\partial K}{\partial p_j}, \quad j = 1,2,
\end{equation}
where
\begin{equation}\label{3.5a}
K = i \left[ \langle\Lambda^2 \textsf{p},\textsf{q} \rangle + \frac{1}{2} \langle\Lambda \textsf{p},\textsf{p} \rangle \langle \textsf{q},\textsf{q} \rangle +
\frac{1}{2} \langle \textsf{p},\textsf{p} \rangle \langle \Lambda
\textsf{q},\textsf{q} \rangle + \frac{1}{2} \langle \textsf{p},\textsf{q}
\rangle \langle \textsf{p},\textsf{p} \rangle \langle \textsf{q},\textsf{q}
\rangle \right].
\end{equation}

There exist four real-valued constants of motion for the Hamiltonian systems (\ref{3.4})--(\ref{3.5})
and (\ref{3.4a})--(\ref{3.5a}),
hence each Hamiltonian system is Liouville integrable. The four constants of motion
were found in \cite{ZhouJMP,ZhouStudies}:
\begin{align}
\phantom{t} & F_0 = i \langle \textsf{p}, \textsf{q} \rangle,
\label{3.9}  \\
\phantom{t} & F_1 = \langle\Lambda \textsf{p}, \textsf{q}\rangle + \frac{1}{2}
\langle \textsf{p},\textsf{p} \rangle \langle \textsf{q}, \textsf{q} \rangle -
\frac{1}{2} \langle \textsf{p},\textsf{q} \rangle^2, \label{3.10} \\
\phantom{t} & F_2 = i \left[ \langle\Lambda^2 \textsf{p},\textsf{q} \rangle +
\frac{1}{2} \langle\Lambda \textsf{p},\textsf{p} \rangle \langle
\textsf{q},\textsf{q} \rangle
+ \frac{1}{2} \langle \textsf{p},\textsf{p} \rangle \langle \Lambda \textsf{q},\textsf{q} \rangle - \langle \textsf{p},\textsf{q} \rangle \langle \Lambda \textsf{p},\textsf{q} \rangle \right],
\label{3.11} \\
\label{3.12}
\phantom{t} & F_3 = \langle \Lambda^3 \textsf{p},\textsf{q} \rangle +
\frac{1}{2} \langle\Lambda^2 \textsf{p},\textsf{p} \rangle \langle
\textsf{q},\textsf{q} \rangle + \frac{1}{2} \langle\Lambda
\textsf{p},\textsf{p} \rangle \langle\Lambda \textsf{q},\textsf{q} \rangle
+ \frac{1}{2} \langle \textsf{p},\textsf{p} \rangle \langle\Lambda^2
\textsf{q}, \textsf{q} \rangle \\
\nonumber
\phantom{t} & \phantom{text} - \frac{1}{2} \langle\Lambda \textsf{p},\textsf{q}
\rangle^2
- \langle \textsf{p},\textsf{q} \rangle \langle\Lambda^2 \textsf{p},\textsf{q}
\rangle.
\end{align}
Note that $H = F_1 - \frac{1}{2} F_0^2$ and $K = F_2 + F_1 F_0 - \frac{1}{2} F_0^3$.

\subsection{Fourth-order Lax--Novikov equation}

In what follows, we relate the $x$-derivatives of $u$ with the four constants of motion $(F_0,F_1,F_2,F_3)$.
By differentiating equation (\ref{3.3}) in $x$ and using the Hamiltonian system
(\ref{3.4})--(\ref{3.5}) together with the first constant (\ref{3.9}), we obtain
\begin{equation}\label{3.13}
\frac{d u}{d x} + 2 i F_0 u = 2 \langle \Lambda \textsf{p},\textsf{p} \rangle.
\end{equation}

By taking another derivative of equation (\ref{3.13}) in $x$ and using the first two constants
(\ref{3.9}) and (\ref{3.10}), we obtain
\begin{equation}\label{3.14}
\frac{d^2 u}{d x^2} + 2 |u|^2 u + 2 i F_0 \frac{du}{dx} - 4 H u
= 4 \langle \Lambda^2 \textsf{p},\textsf{p} \rangle,
\end{equation}
where $H = F_1 - \frac{1}{2} F_0^2$ is the Hamiltonian for system (\ref{3.4}).

By taking another derivative of equation (\ref{3.14}) in $x$ and using the first three constants
(\ref{3.9})-(\ref{3.11}), we obtain
\begin{equation}\label{3.15}
\begin{split}
&\frac{d^3 u}{d x^3} + 6 |u|^2 \frac{du}{dx} + 2 i F_0 \left( \frac{d^2 u}{d
x^2} + 2 |u|^2 u \right) \\
&- 4 H \frac{du}{dx} + 8 i K u = 8 \langle\Lambda^3 \textsf{p},\textsf{p}
\rangle,
\end{split}
\end{equation}
where $K = F_2 + F_0 H = F_2 + F_0 F_1 - \frac{1}{2} F_0^3$ is the Hamiltonian for system (\ref{3.4a}).

Finally, by taking yet another derivative of equation (\ref{3.15}) in $x$ and using all four constants
(\ref{3.9})-(\ref{3.12}), we obtain
\begin{eqnarray}
\label{3.16}
&& \frac{d^4 u}{d x^4} + 8 |u|^2 \frac{d^2 u}{dx^2} + 2 u^2 \frac{d^2 \bar{u}}{d x^2}
+ 4 u \left| \frac{du}{dx} \right|^2 + 6 \left(\frac{du}{dx}\right)^2 \bar{u}
+ 6 |u|^4 u + 2 i F_0 \left( \frac{d^3 u}{d x^3} + 6 |u|^2 \frac{du}{dx} \right) \\
\nonumber
&&
\quad - 4H \left( \frac{d^2 u}{dx^2} + 2 |u|^2 u \right) + 8 i K \frac{du}{dx}
- 16 E u = 16\langle\Lambda^4 \textsf{p},\textsf{p} \rangle,
\end{eqnarray}
where $E = F_3 - F_0 K + \frac{1}{2} H^2 =
F_3 - F_0 F_2 + \frac{1}{2} F_1^2 - \frac{3}{2} F_1 F_0^2 + \frac{5}{8} F_0^4$.

The system of linear algebraic equations on squared eigenfunctions
$p_1^2$, $p_2^2$, $\bar{q}_1^2$, and $\bar{q}_2^2$ is fully determined
by four equations (\ref{3.3}), (\ref{3.13}), (\ref{3.14}), and (\ref{3.15}).
Therefore, the fourth-order equation (\ref{3.16}) can be closed on $u$
as the fourth-order Lax--Novikov equation for the hierarchy of stationary NLS equations.
In order to avoid the linear algebra computations, we will use integrability
of the Hamiltonian system (\ref{3.4})--(\ref{3.5}) and obtain the closed
fourth-order equation on $u$ by a simple computation.

\subsection{Integrability of the fourth-order Lax--Novikov equation}

It was shown in \cite{ZhouJMP,ZhouStudies} that the Hamiltonian system (\ref{3.4})--(\ref{3.5})
arises as a compatibility condition for the Lax equation
\begin{equation}\label{3.7}
\frac{d}{dx} W(\lambda) = [U(\lambda,u),W(\lambda)], \qquad \lambda \in
\mathbb{C},
\end{equation}
where $U(\lambda,u)$ is given by (\ref{3.1}) with $u$ given by (\ref{3.3}) and
\begin{equation}\label{3.6}
\displaystyle
W(\lambda) =\left(\begin{array}{cc}
W_{11}(\lambda) & W_{12}(\lambda) \\
\overline{W}_{12}(-\lambda) & -\overline{W}_{11}(-\lambda) \end{array}\right),
\end{equation}
with the entries
\begin{equation}
\label{W-11-12}
W_{11}(\lambda) =1-\sum\limits_{j=1}^2\left(\frac{p_jq_j}{\lambda-\lambda_j}-\frac{\bar{p}_j\bar{q}_j}{\lambda+\bar{\lambda}_j}\right), \quad
W_{12}(\lambda)=\sum\limits_{j=1}^2\left(\frac{p_j^2}{\lambda-\lambda_j}+\frac{\bar{q}^2_j}{\lambda+\bar{\lambda}_j}\right).
\end{equation}
In order to progress further, we rewrite $W_{12}(\lambda)$ in the equivalent form:
\begin{equation}\label{3.19}
W_{12}(\lambda) = \frac{S_0 \lambda^3 + S_1 \lambda^2 + S_2 \lambda +
S_3}{\lambda^4 + i A_1 \lambda^3 + A_2 \lambda^2 + i A_3 \lambda + A_4},
\end{equation}
where
\begin{align}
\left\{ \begin{array}{l}
A_1 = i(\lambda_1+\lambda_2 -\bar{\lambda}_1-\bar{\lambda}_2),\\
A_2 = (\lambda_1-\bar{\lambda}_1)(\lambda_2-\bar{\lambda}_2)
-|\lambda_1|^2-|\lambda_2|^2,\\
A_3 = i(\bar{\lambda}_1 - \lambda_1) |\lambda_2|^2 + i(\bar{\lambda}_2 -
\lambda_2) |\lambda_1|^2,\\
A_4 = |\lambda_1|^2|\lambda_2|^2, \end{array} \right.
\label{A-coefficients}
\end{align}
and
\begin{align}
\left\{ \begin{array}{l}
S_0 = \langle \textsf{p}, \textsf{p} \rangle, \\
S_1 = \langle \Lambda \textsf{p},\textsf{p} \rangle + i A_1 \langle \textsf{p},\textsf{p}\rangle, \\
S_2 = \langle \Lambda^2 \textsf{p},\textsf{p} \rangle + i A_1 \langle \Lambda \textsf{p},\textsf{p}\rangle + A_2\langle \textsf{p},\textsf{p}\rangle,\\
S_3 = \langle \Lambda^3 \textsf{p},\textsf{p}\rangle + i A_1 \langle \Lambda^2 \textsf{p},\textsf{p}\rangle + A_2 \langle \Lambda \textsf{p},\textsf{p}\rangle + i A_3 \langle \textsf{p},\textsf{p}\rangle.
\end{array} \right.
\label{S-expressions}
\end{align}
Substituting (\ref{3.3}), (\ref{3.13}), (\ref{3.14}), and (\ref{3.15}) into these expressions yield
compact expressions:
\begin{equation}
\label{3.20}
\left\{
\begin{array}{l}
S_0 = u, \\
S_1 = \frac{1}{2} u' + i c u,\\
S_2 = \frac{1}{4} (u'' + 2 |u|^2 u) +\frac{i}{2} c u' + b u,\\
S_3 = \frac{1}{8} (u''' + 6 |u|^2 u') + \frac{i}{4} c (u'' + 2|u|^2 u) + \frac{1}{2} b u' + i a u,
\end{array} \right.
\end{equation}
where the prime denotes the derivative in $x$ and we have introduced three real-valued constants:
\begin{equation}
\label{3.20a}
\left\{
\begin{array}{l}
c = A_1 + F_0, \\
b = A_2 - F_0 A_1 - H,\\
a = A_3 + F_0 A_2 - H A_1 + K,
\end{array} \right.
\end{equation}
Similarly, we rewrite $W_{11}(\lambda)$ in the equivalent form:
\begin{equation}\label{3.17}
W_{11}(\lambda) = \frac{\lambda^4 + i T_1 \lambda^3 + T_2 \lambda^2 + i T_3 \lambda + T_4}{
\lambda^4 + i A_1 \lambda^3 + A_2 \lambda^2 + i A_3 \lambda + A_4},
\end{equation}
where
\begin{align}
\left\{ \begin{array}{l}
T_1 = A_1 + i \langle \textsf{p},\textsf{q}\rangle, \\
T_2 = A_2 - i A_1 \langle \textsf{p},\textsf{q}\rangle - \langle \Lambda \textsf{p},\textsf{q}\rangle, \\
T_3 = A_3 + i A_2 \langle \textsf{p},\textsf{q}\rangle - A_1 \langle \Lambda \textsf{p},\textsf{q}\rangle + i \langle \Lambda^2 \textsf{p},\textsf{q}\rangle, \\
T_4 = A_4 - i A_3 \langle \textsf{p},\textsf{q}\rangle - A_2 \langle \Lambda \textsf{p},\textsf{q}\rangle - i A_1 \langle \Lambda^2 \textsf{p},\textsf{q}\rangle
- \langle \Lambda^3 \textsf{p},\textsf{q}\rangle.
\end{array} \right.
\label{T-expressions}
\end{align}
Substituting (\ref{3.9}), (\ref{3.10}), (\ref{3.11}), and (\ref{3.12}) into these expressions yield
compact expressions:
\begin{equation}
\label{3.18}
\left\{
\begin{array}{l}
T_1 = c, \\
T_2 = b + \frac{1}{2} |u|^2, \\
T_3 = a + \frac{1}{2} c |u|^2 - \frac{i}{4} (u' \bar{u}-u \bar{u}'),\\
T_4 = d + \frac{1}{2} b |u|^2 + \frac{i}{4} c (u' \bar{u}-u \bar{u}')
+ \frac{1}{8} (u \bar{u}'' + u'' \bar{u} - |u'|^2 + 3 |u|^4),
\end{array} \right.
\end{equation}
where we have used (\ref{3.20a}) and introduced another real-valued constant:
\begin{equation}
\label{3.20b}
d = A_4 - F_0 A_3 - H A_2 - K A_1 - E.
\end{equation}
The $(1,2)$-component of the Lax equation (\ref{3.7}) is written explicitly in the form:
\begin{equation}\label{3.21}
\frac{d}{dx} W_{12}(\lambda) = 2 \lambda W_{12}(\lambda) - 2 \langle
\textsf{p},\textsf{p}\rangle W_{11}(\lambda),
\end{equation}
Substituting (\ref{3.19}) and (\ref{3.17})
with (\ref{3.20}) and (\ref{3.18}) into (\ref{3.21}) yields constraints
at different powers of $\lambda$. However, all constraints are satisfied
identically at powers of $\lambda^3$, $\lambda^2$ and $\lambda$,
whereas the constraint at $\lambda^0$, that is, $\frac{d}{dx} S_3 + 2u T_4 = 0$, is equivalent to the fourth-order
Lax--Novikov equation:
\begin{align}
\label{3.22}
& u'''' + 8 |u|^2 u'' + 2 u^2 \bar{u}''
+ 4 u |u'|^2 + 6 (u')^2 \bar{u}
+ 6 |u|^4 u \\
& + 2 i c (u''' + 6 |u|^2 u') + 4b (u'' + 2 |u|^2 u) + 8ia u' + 16 d u = 0,
\nonumber
\end{align}
which provides a closed fourth-order equation on $u$ compared to (\ref{3.16}).

The fourth-order complex-valued Lax--Novikov equation (\ref{3.22}) is integrable with two complex-valued constants of motion.
In order to derive them, we recall from \cite{ZhouJMP,ZhouStudies} that
the determinant of $W(\lambda)$ is a constant of motion and has only simple poles at
$\lambda_1$, $\lambda_2$, $-\bar{\lambda}_1$, and $-\bar{\lambda}_2$.
On the other hand, it follows from (\ref{3.6}), (\ref{3.19}), and (\ref{3.17}) that
\begin{align}
\label{det-W}
\quad \det W(\lambda) = - \left[ W_{11}(\lambda) \right]^2 - W_{12}(\lambda) \bar{W}_{12}(-\lambda)
= - \frac{P(\lambda)}{(\lambda-\lambda_1)^2 (\lambda-\lambda_2)^2
(\lambda+\bar{\lambda}_1)^2 (\lambda+\bar{\lambda}_2)^2},
\end{align}
where
\begin{equation}\label{3.24}
\quad P(\lambda) := (\lambda^4 + i T_1\lambda^3 + T_2\lambda^2 + i T_3\lambda + T_4)^2 -
(S_0 \lambda^3 + S_1 \lambda^2 + S_2 \lambda+ S_3)
(\bar{S}_0 \lambda^3 - \bar{S}_1 \lambda^2 + \bar{S}_2 \lambda - \bar{S}_3).
\end{equation}
Since $\lambda_1$, $\lambda_2$, $-\bar{\lambda}_1$, and $-\bar{\lambda}_2$ are roots of $P(\lambda)$,
substituting (\ref{3.20}) and (\ref{3.18}) into (\ref{3.24}) and evaluating at $\lambda_1$ and $\lambda_2$ yield
two complex-valued constants of motion for the fourth-order Lax--Novikov equation (\ref{3.22}).
The following symmetry of roots of $P(\lambda)$ provides complex-conjugate symmetry of the two
complex-valued constants of motion: {\em If $\lambda_0$ is a root of $P(\lambda)$, so is $-\bar{\lambda}_0$,
thanks to the symmetry of the coefficients in $P(\lambda)$}. As a result, ${\rm det} W(\lambda)$ has simple poles
at $\lambda_1$, $\lambda_2$, $-\bar{\lambda}_1$, and $-\bar{\lambda}_2$ in the quotient given by (\ref{det-W}).

The admissible values of the algebraic method for $\lambda_1$ and $\lambda_2$ are defined
from the four pairs of roots of $P(\lambda)$ which are symmetric about the imaginary axis.
Therefore, we are free to choose $\lambda_1$ and $\lambda_2$ from any of the four pairs of roots of $P(\lambda)$.
In Section \ref{section-3}, we adopt the algebraic method with two eigenvalues to recover the double-periodic solutions
(\ref{solB}) and (\ref{solA}).

\section{Double-periodic waves}
\label{section-3}

Truncation of polynomials for $W_{11}(\lambda)$ and $W_{12}(\lambda)$
by setting $S_3 = 0$ and $T_4 = 0$  yield the third-order Lax--Novikov equation
studied in \cite{Wright}. Since the third-order Lax--Novikov equation
contains all solutions written in terms of the Riemann Theta function of genus two
including the double-periodic solutions (\ref{solB}) and (\ref{solA}), we restrict our work to exploring this reduction.
Setting $S_3 = 0$ and $T_4 = 0$ yields the third-order differential equation
\begin{equation}
\label{LN-3}
u''' + 6 |u|^2 u' + 2i c (u'' + 2|u|^2 u) + 4 b u' + 8i a u = 0,
\end{equation}
and its second-order invariant
\begin{equation}
\label{Constant-3a}
d + \frac{1}{2} b |u|^2 + \frac{i}{4} c (u' \bar{u}-u \bar{u}')
+ \frac{1}{8} (u \bar{u}'' + u'' \bar{u} - |u'|^2 + 3 |u|^4) = 0.
\end{equation}
The fourth-order equation (\ref{3.22}) is now satisfied identically in view of
(\ref{LN-3}) and (\ref{Constant-3a}).

Since $P(\lambda)$ is independent on $x$, substituting (\ref{3.20}) and (\ref{3.18}) into
(\ref{3.24}) and expanding in $\lambda$ yields two more invariants in powers $\lambda$ and $\lambda^0$:
\begin{equation}
\label{Constant-3b}
2e - a |u|^2 - \frac{1}{4} c (|u'|^2 + |u|^4)
+ \frac{i}{8} (u'' \bar{u}' - u' \bar{u}'') = 0,
\end{equation}
and
\begin{equation}
\label{Constant-3c}
f - \frac{i}{2} a (u' \bar{u}-u \bar{u}') + \frac{1}{4} b (|u'|^2 + |u|^4)
+ \frac{1}{16} (|u'' + 2 |u|^2 u|^2 - (u' \bar{u} - u \bar{u}')^2) = 0,
\end{equation}
where $(e,f)$ are additional real-valued constants to the previous list $(a,b,c,d)$ defined in
(\ref{3.20a}) and (\ref{3.20b}). It has been verified directly that (\ref{Constant-3a}),
(\ref{Constant-3b}) and (\ref{Constant-3c}) are constants of motion for the third-order equation (\ref{LN-3}).

With the account of (\ref{LN-3}), (\ref{Constant-3a}), (\ref{Constant-3b}) and (\ref{Constant-3c}),
it follows from (\ref{3.24}) that $P(\lambda) = \lambda^2 \tilde{P}(\lambda)$, where
\begin{align}
\label{Polynomial-3}
&\tilde{P}(\lambda)  := \lambda^6 + 2 i c \lambda^5 + (2b-c^2) \lambda^4 + 2i
(a + bc) \lambda^3 +
(b^2 - 2 ac + 2 d) \lambda^2 \\
\nonumber
 \phantom{t}  \phantom{text} &+ 2i (e + ab + cd) \lambda + f + 2bd - 2ce - a^2.
\end{align}
There exist three admissible pairs of eigenvalues found from roots of $\tilde{P}(\lambda)$, two of which
can be taken as $\lambda_1$ and $\lambda_2$. In order to express
the three pairs of roots of $\tilde{P}(\lambda)$ explicitly in terms of parameters $(a,b,c,d,e,f)$,
we reduce the third-order Lax--Novikov equation (\ref{LN-3}) to the first-order quadratures.
This requires us to set $a = c = e = 0$, whereas the other three constants $(b,d,f)$ are left arbitrarily.

\subsection{Reduction to the first-order quadratures}

The following transformation
\begin{align}
u(x) = \tilde{u}(x) e^{-\frac{2}{3} i c x}
\end{align}
with
\begin{eqnarray*}
b = \tilde{b} - \frac{1}{3} c^2, \quad
a = \tilde{a} + \frac{1}{3} bc + \frac{2}{27} c^3, \quad
d = \tilde{d}, \quad
e = \tilde{e} - \frac{1}{3} c d, \quad
f = \tilde{f} + \frac{4}{3} ce + \frac{2}{9} c^2 d
\end{eqnarray*}
leaves the system (\ref{LN-3}), (\ref{Constant-3a}), (\ref{Constant-3b}), and (\ref{Constant-3c})
invariant for tilde variables and eliminates the parameter $c$. Hence, $c = 0$ can be set without loss of generality.
In addition, we set $a = e = 0$ in order to reduce
the system (\ref{LN-3}), (\ref{Constant-3a}), (\ref{Constant-3b}), and (\ref{Constant-3c})
to the first-order quadratures.

It follows from (\ref{Constant-3b}) with $a = c = e = 0$ that
\begin{equation}
\frac{d}{dx} \log\left( \frac{u'}{\bar{u}'}\right) = 0 \quad \Rightarrow \quad
\frac{u'}{\bar{u}'} = e^{2i\theta},
\end{equation}
where real $\theta$ is constant in $x$. Hence, $u'(x) = e^{i\theta} Q'(x)$ with real $Q$
and integrating it again and adding the time variable $t$, we obtain the
following form for the solution to the NLS equation (\ref{nls}):
\begin{equation}
\label{double-periodic}
\psi(x,t) = \left[ Q(x,t) + i \delta(t) \right] e^{i \theta(t)},
\end{equation}
where real $\delta$ is constant in $x$. The exact solution to the NLS equation (\ref{nls})
in the form (\ref{double-periodic}) was characterized in \cite{Nail1} by
using separation of variables. Here we characterize the same solution
by using the third-order Lax--Novikov equation (\ref{LN-3})
with the remaining two conserved quantities in (\ref{Constant-3a}) and (\ref{Constant-3c}). Substituting
\begin{equation}
\label{double-periodic-u}
u = (Q + i \delta) e^{i \theta},
\end{equation}
with $x$-independent $\delta$ and $\theta$ into
the third-order equation (\ref{LN-3}) with $a = c = 0$ yields
\begin{equation}
Q_{xxx} + 6 (Q^2 + \delta^2) Q_x + 4 b Q_x = 0.
\end{equation}
Integrating this third-order equation in $x$ yields the following second-order equation:
\begin{equation}
\label{Q-equation}
Q_{xx} + 2 Q^3 + 6 \delta^2 Q + 4b Q = A,
\end{equation}
where $A$ is independent of $x$ but may depend on $t$. Substituting (\ref{double-periodic-u}) and (\ref{Q-equation})
into (\ref{Constant-3a}) with $c = 0$ yields the first-order quadrature:
\begin{equation}
\label{Q-integral}
\left(\frac{dQ}{dx} \right)^2 + F(Q) = 0, \quad F(Q) = Q^4 + (6 \delta^2 + 4b)
Q^2 - 2 A Q - 3 \delta^4 - 4b \delta^2 - 8d.
\end{equation}
Substituting (\ref{double-periodic-u}), (\ref{Q-equation}), and (\ref{Q-integral}) into
(\ref{Constant-3c}) with $a = c = 0$ yields
\begin{equation}
\label{A-equation}
A^2 + 16 \left[ (\delta^2 + b)( \delta^4 + b \delta^2 + 2 d) + f \right] = 0.
\end{equation}
It remains to relate $A$ with $\delta$, for which we use the time evolution
of the NLS equation (\ref{nls}). Substituting (\ref{double-periodic}) into (\ref{nls})
and separating the variables yield the following system:
\begin{align}
\label{Q-system}
\left. \begin{array}{r}
Q_{xx} + 2(Q^2 + \delta^2 - \dot{\theta}) Q - 2 \dot{\delta} = 0, \\
Q_{t} + (Q^2 + \delta^2 - \dot{\theta}) \delta = 0, \end{array} \right\}
\end{align}
where the dot denotes the derivative of $\delta$ and $\theta$ in $t$.
Comparing (\ref{Q-equation}) with the first equation of system (\ref{Q-system})
yields $A = 2 \dot{\delta}$ and $\dot{\theta} = - 2 (\delta^2 + b)$.
Setting $z := \delta^2$ reduces (\ref{A-equation}) with $A = 2 \dot{\delta}$ to the first-order quadrature:
\begin{equation}
\label{Jacobi-z}
\left( \frac{d z}{dt} \right)^2 + G(z) = 0, \quad
G(z) = 16 z (z^3 + 2 b z^2 + (b^2 + 2 d) z + f + 2 b d).
\end{equation}
Let us parameterize the constant $b$, $d$, and $f$ as follows:
\begin{equation}
\label{roots-constants}
\left\{ \begin{array}{l}
-2b = z_1 + z_2 +z_3, \\
2d + b^2 = z_1 z_2 + z_1 z_3 + z_2 z_3, \\
f + 2bd = -z_1 z_2 z_3.
\end{array} \right.
\end{equation}
In this case, the quadratures in (\ref{Q-integral}) and (\ref{Jacobi-z}) are parameterized by
\begin{equation}
\label{expression-F}
F(Q) = Q^4 + 2( 3z - z_1 - z_2 - z_3) Q^2 - 4 \dot{\delta} Q - 3z^2 + 2z (z_1 +
z_2 + z_3) - 2(z_1 z_2 + z_1 z_3 + z_2 z_3) + z_1^2 + z_2^2 + z_3^2
\end{equation}
and
\begin{equation}
\label{expression-G}
G(z) = 16 z (z-z_1) (z - z_2) (z-z_3).
\end{equation}
The polynomial $\tilde{P}(\lambda)$ in (\ref{Polynomial-3})
is transformed under the parametrization (\ref{roots-constants})
with $a = c = e = 0$ to the form:
\begin{align}
\tilde{P}(\lambda) = \lambda^6 - (z_1+z_2+z_3) \lambda^4 +
(z_1z_2 + z_1 z_3 + z_2 z_3) \lambda^2 - z_1z_2z_3,
\label{Polynomial-3-explicit}
\end{align}
which shows that $\pm \sqrt{z_1}$, $\pm \sqrt{z_2}$, $\pm \sqrt{z_3}$ are roots of
$\tilde{P}(\lambda)$.

\subsection{Exact solutions with elliptic functions}

We obtain the exact solutions of the first-order quadratures (\ref{Q-integral}) and (\ref{Jacobi-z})
with $F(Q)$ and $G(z)$ given by (\ref{expression-F}) and (\ref{expression-G}). Since $z = \delta^2 \geq 0$, one of the roots
$z_{1,2,3}$ must be positive, whereas the other two roots are either real or complex-conjugate.

\subsubsection{Real roots}

When the three roots $z_{1,2,3}$ are real, let us order them by $z_1 \leq z_2 \leq z_3$. From positivity
of the solutions $z(t)$ of the first-order quadrature (\ref{Jacobi-z}) with (\ref{expression-G}) it follows
that $z_3 > 0$ and $z(t) \in [0,z_3]$. Proceeding as in \cite{Nail1}, we
replace formally $\dot{\delta} = 2 \sqrt{(z_1-z)(z_2-z)(z_3-z)}$ and factorize
the quartic polynomial $F(Q)$ in (\ref{expression-F}) for $z \in [0,z_3]$ by
\begin{align}
\label{expression-F-real}
&F(Q)  =  \left[ Q^2 + 2 \sqrt{z_3 - z} Q + z_3 + z - z_1 - z_2 + 2
\sqrt{(z_1-z)(z_2-z)} \right] \\ \nonumber
 \phantom{t}  \times
&\left[ Q^2 - 2 \sqrt{z_3 - z} Q + z_3 + z - z_1 - z_2 - 2
\sqrt{(z_1-z)(z_2-z)}
\right],
\end{align}
so that the discriminants of the two quadratic equations are given by
\begin{equation}
\label{discriminents}
D_{\pm} = z_1 + z_2 - 2z \pm 2 \sqrt{(z_1-z)(z_2-z)}.
\end{equation}
It follows from the discriminants (\ref{discriminents}) that all four roots of $F(Q)$ are complex-valued
unless $z \leq z_1$. Hence $z_1 > 0$ and $z(t) \in [0,z_1]$ so that the three real roots satisfy the following ordering:
\begin{equation}
\label{z-roots}
0 \leq z_1 \leq z_2 \leq z_3.
\end{equation}
The exact solution of the quadrature (\ref{Jacobi-z}) for $z(t) \in [0,z_1]$ under the ordering (\ref{z-roots}) is given by
the following explicit expression:
\begin{equation}
\label{Jacobi-1-z}
z(t) = \frac{z_1 z_3 {\rm sn}^2(\mu t;k)}{z_3 - z_1 {\rm cn}^2(\mu t; k)},
\end{equation}
where $\mu$ and $k$ are related to $(z_1,z_2,z_3)$ by
$$
\mu^2 = 4 z_2 (z_3 - z_1), \quad k^2 = \frac{z_1(z_3-z_2)}{z_2(z_3-z_1)}.
$$
The validity of the explicit expression (\ref{Jacobi-1-z}) can be verified
from (\ref{Jacobi-z}) and (\ref{expression-G}) by explicit substitution, see \cite{CPgardner} for details.
Extracting the square root from $z = \delta^2$ in such a way that $\delta(t)$ remains smooth in $t$ yields the exact solution:
\begin{equation}
\label{Jacobi-1-delta}
\delta(t) = \frac{\sqrt{z_1 z_3} {\rm sn}(\mu t;k)}{\sqrt{z_3 - z_1 {\rm
cn}^2(\mu t; k)}},
\end{equation}
which is a smooth periodic function of $t$ with period $T = 4 K(k)/\mu$.

When $z(t) \in [0,z_1]$ under the ordering (\ref{z-roots}), the four roots of $F(Q)$ in
the factorization (\ref{expression-F-real}) are real and can be ordered as
\begin{equation}
\label{Q-roots-order}
Q_4 \leq Q_3 \leq Q_2 \leq Q_1.
\end{equation}
The following explicit formula for the roots $Q_{1,2,3,4}$ was used in \cite{Nail1}:
\begin{align}
\label{Q-roots}
\left\{ \begin{array}{l}
Q_1 = \sqrt{z_1 - z} + \sqrt{z_2 - z} + \sqrt{z_3-z}, \\
Q_2 = -\sqrt{z_1 - z} - \sqrt{z_2 - z} + \sqrt{z_3-z}, \\
Q_3 = -\sqrt{z_1 - z} + \sqrt{z_2 - z} - \sqrt{z_3-z}, \\
Q_4 = \sqrt{z_1 - z} - \sqrt{z_2 - z} - \sqrt{z_3-z},
\end{array} \right. 	
\end{align}
and can be verified by the explicit computations from (\ref{expression-F-real}).
However, since $\sqrt{z_1 - z(t)}$ is non-smooth at $t = K(k)/\mu$, this parametrization
introduces singularities in the definition of $Q(x,t)$.
Similarly, one can use the parametrization (\ref{Q-roots}) with $\sqrt{z_1 - z}$ replaced by $-\sqrt{z_1 - z}$
but this choice also introduces singularities in the definition of $Q(x,t)$.

In order to avoid the branch point singularities, we shall parameterize the roots
$Q_{1,2,3,4}$ by using the original representation (\ref{expression-F}) in the form:
\begin{align}
\label{Q-roots-correct}
\left\{ \begin{array}{l}
Q_1 = \sqrt{z_3-z} + \sqrt{z_1 + z_2 - 2 z + \dot{\delta}/\sqrt{z_3 - z}}, \\
Q_2 = \sqrt{z_3-z} - \sqrt{z_1 + z_2 - 2 z + \dot{\delta}/\sqrt{z_3 - z}}, \\
Q_3 = -\sqrt{z_3-z} + \sqrt{z_1 + z_2 - 2 z - \dot{\delta}/\sqrt{z_3 - z}}, \\
Q_4 = -\sqrt{z_3-z} - \sqrt{z_1 + z_2 - 2 z - \dot{\delta}/\sqrt{z_3 - z}},
\end{array} \right.
\end{align}
where all square roots stay away from zero, so that the definition of $Q(x,t)$ is smooth.

The exact solution of the quadrature (\ref{Q-integral}) with
$$
F(Q) = (Q-Q_1) (Q-Q_2) (Q-Q_3) (Q - Q_4)
$$
for $Q(x,t) \in [Q_2,Q_1]$ is given by
\begin{equation}
Q(x,t) = Q_4 + \frac{(Q_1-Q_4) (Q_2-Q_4)}{(Q_2-Q_4) + (Q_1 - Q_2) {\rm sn}^2(\nu x;\kappa)},
\label{Jacobi-1}
\end{equation}
where
$$
\left\{ \begin{array}{l}
4 \nu^2 = (Q_1 - Q_3) (Q_2 - Q_4), \\
4 \nu^2 \kappa^2 = (Q_1 - Q_2) (Q_3 - Q_4), \end{array} \right.
\quad \Rightarrow \quad
\left\{ \begin{array}{l}
\nu^2 = z_3-z_1, \\
\nu^2 \kappa^2 = z_2-z_1. \end{array} \right.
$$
The function $Q$ is periodic in $x$ with period $L = 2K(\kappa)/\nu$
and periodic in $t$ with period $T = 4 K(k)/\mu$ thanks to the $T$-periodicity
of $z(t)$ and $\dot{\delta}(t)$ in (\ref{Q-roots-correct}). Similarly, the exact solution
of the quadrature (\ref{Q-integral}) for $Q(x,t) \in [Q_4,Q_3]$
is given by
\begin{equation}
Q(x,t) = Q_2 + \frac{(Q_3-Q_2) (Q_4-Q_2)}{(Q_4-Q_2) + (Q_3 - Q_4) {\rm sn}^2(\nu x;\kappa)},
\label{Jacobi-1-symm}
\end{equation}
with the same values for $\nu$ and $\kappa$. Solution (\ref{Jacobi-1}) and (\ref{Jacobi-1-symm}) have the same periodicity
in $x$ and $t$.

At $t = 0$, we have $\delta(0) = 0$, whereas $\theta(0) = 0$ can be chosen without loss of generality,
thanks to the gauge transformation of the NLS equation (\ref{nls}).
In this case, the real-valued function $\psi(x,0) = Q(x,0)$ in (\ref{double-periodic})
coincide with the general periodic wave solution of the modified KdV equation
considered in \cite{CPgardner}. The roots of the polynomial $\tilde{P}(\lambda)$ in (\ref{Polynomial-3-explicit})
can then be written in the form:
\small\begin{equation}
\label{eig-Q-real}
\lambda_1^{\pm} = \pm \sqrt{z_1} = \pm \frac{Q_1 + Q_4}{2} \biggr|_{t = 0},\;\;
\lambda_2^{\pm} = \pm \sqrt{z_2} = \pm \frac{Q_1 + Q_3}{2} \biggr|_{t = 0},\;\;
\lambda_3^{\pm} = \pm \sqrt{z_3} = \pm \frac{Q_1 + Q_2}{2} \biggr|_{t = 0}.
\end{equation}
For $t \neq 0$, the complex-valued function $\psi(x,t)$ in (\ref{double-periodic}) describes a double-periodic
solution to the NLS equation (\ref{nls}). In the special case $z_3 = z_1 + z_2 = 1$,
the exact solution can be simplified to the form (\ref{solB}) derived in \cite{Nail1}.
Appendix A gives technical details of the relevant transformations.
\begin{figure}[h!]
	\centering
	\includegraphics[width=8.5cm,height=6.5cm]{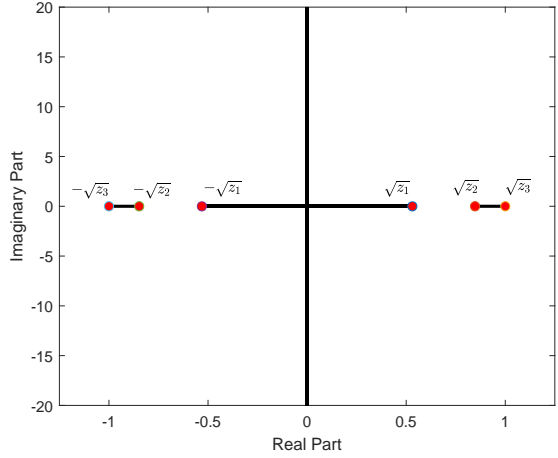}
	\caption{Lax spectrum for the double-periodic solution (\ref{solB}) with
	$k=0.9$. Red dots represent eigenvalues $\pm \sqrt{z_1}$, $\pm \sqrt{z_2}$, and $\pm \sqrt{z_3}$.}
	\label{f6}
\end{figure}

Figure \ref{f6} represents the Lax spectrum
computed numerically by using the Floquet--Bloch decomposition
of solutions to the spectral problem (\ref{3.1}) with the periodic potential $u$
in the form (\ref{solB}). The numerical method is described in Appendix A in our
previous work \cite{CPWnls}. The black curves represent the purely continuous spectrum whereas
the red dots represent eigenvalues (\ref{eig-Q-real}).

\subsubsection{Complex-conjugate roots}

When one root in $z_{1,2,3}$ is real and positive while the other two roots are
complex-conjugate, we can introduce the parametrization:
\begin{equation}
\label{z-roots-complex}
z_1 \geq 0, \quad z_2 = \xi + i \eta, \quad z_3 = \xi - i \eta
\end{equation}
with real-valued $\xi$ and $\eta$. The exact solution of the quadrature (\ref{Jacobi-z}) for $z(t) \in [0,z_1]$ under the parametrization (\ref{z-roots-complex})
is given by
\begin{equation}
\label{Jacobi-2-z}
z(t) = \frac{z_1 (1 - {\rm cn}(\mu t; k))}{(1 + \zeta) + (\zeta - 1) {\rm
cn}(\mu t;k)},
\end{equation}
where $\zeta > 0$, $\mu > 0$, and $k \in (0,1)$ are parameters given by
$$
\zeta = \frac{\sqrt{(z_1-\xi)^2 + \eta^2}}{\sqrt{\xi^2 + \eta^2}}, \quad
\mu^2 = 16 \sqrt{(z_1-\xi)^2 + \eta^2} \sqrt{\xi^2 + \eta^2}, \quad
2 k^2 = 1 + \frac{\xi (z_1-\xi) - \eta^2}{\sqrt{(z_1-\xi)^2 + \eta^2} \sqrt{\xi^2 + \eta^2}}.
$$
We need to extract the square root from $z = \delta^2$ so that $\delta(t)$ remain smooth
in $t$. To do so, we use the half-argument formula
$$
{\rm sn}^2\left(\frac{1}{2} \mu t;k \right) = \frac{1 - {\rm cn}(\mu t; k)}{1 + {\rm dn}(\mu t;k)},
$$
which yields the exact solution:
\begin{equation}
\label{Jacobi-2-delta}
\delta(t) = \frac{\sqrt{z_1} \sqrt{1 + {\rm dn}(\mu t;k)}}{\sqrt{(1 + \zeta) + (\zeta - 1) {\rm cn}(\mu t;k)}}
{\rm sn}\left(\frac{1}{2} \mu t;k \right),
\end{equation}
which is a smooth periodic function of $t$ with period $T = 8 K(k)/\mu$.

Since $z_1 \geq 0$ is now the largest positive root of $G(z)$, we can rewrite
the factorization (\ref{expression-F-real}) in the equivalent form:
\begin{align}
\label{expression-F-complex}
F(Q) & = \left[ Q^2 + 2 \sqrt{z_1 - z} Q + z_1 + z - 2 \xi + 2 \sqrt{(\xi -
z)^2 + \eta^2} \right] \\
\nonumber
\phantom{t} & \times
\left[ Q^2 - 2 \sqrt{z_1 - z} Q + z_1 + z - 2 \xi - 2 \sqrt{(\xi - z)^2 +
\eta^2} \right],
\end{align}
so that the discriminants of the two quadratic equations are given by
\begin{equation}
\label{discriminents-complex}
D_{\pm} = 2 \xi - 2z \pm 2 \sqrt{(\xi - z)^2 + \eta^2}.
\end{equation}
Note that the factorization formula (\ref{expression-F-complex}) is formal since the sign
of $\sqrt{z_1 - z}$ should be defined from smooth continuations of the roots of $F(Q)$.
It follows from the discriminants (\ref{discriminents-complex}) that two roots of $F(Q)$
are real-valued, ordered as $Q_2 \leq Q_1$, and two roots are complex-conjugate, $Q_3 = \bar{Q}_4$.

The following explicit formula for the roots $Q_{1,2,3,4}$ was used in \cite{Nail1}:
\begin{align}
\label{Q-roots-complex}
\left\{ \begin{array}{l}
Q_1 = \sqrt{z_1-z} + \sqrt{2} \sqrt{\sqrt{(\xi - z)^2 + \eta^2} + (\xi - z)}, \\
Q_2 = \sqrt{z_1-z} - \sqrt{2} \sqrt{\sqrt{(\xi - z)^2 + \eta^2} + (\xi - z)}, \\
Q_3 = -\sqrt{z_1-z} + i \sqrt{2} \sqrt{\sqrt{(\xi - z)^2 + \eta^2} - (\xi - z)}, \\
Q_4 = -\sqrt{z_1-z} - i \sqrt{2} \sqrt{\sqrt{(\xi - z)^2 + \eta^2} - (\xi - z)},
\end{array} \right.
\end{align}
however, $\sqrt{z_1 - z(t)}$ is non-smooth at $t = 2 K(k)/\mu$, which introduces
singularities in the definition of $Q(x,t)$. In order to avoid the branch point singularities,
we replace (\ref{Q-roots-complex}) by
\begin{align}
\label{Q-roots-complex-correct}
\left\{ \begin{array}{l}
Q_1 = \frac{\dot{\delta}}{2 \sqrt{(\xi-z)^2 + \eta^2}} + \sqrt{2} \sqrt{\sqrt{(\xi - z)^2 + \eta^2} + (\xi - z)}, \\
Q_2 = \frac{\dot{\delta}}{2 \sqrt{(\xi-z)^2 + \eta^2}} - \sqrt{2} \sqrt{\sqrt{(\xi - z)^2 + \eta^2} + (\xi - z)}, \\
Q_3 = -\frac{\dot{\delta}}{2 \sqrt{(\xi-z)^2 + \eta^2}} + i \sqrt{2} \sqrt{\sqrt{(\xi - z)^2 + \eta^2} - (\xi - z)}, \\
Q_4 = -\frac{\dot{\delta}}{2 \sqrt{(\xi-z)^2 + \eta^2}} - i \sqrt{2} \sqrt{\sqrt{(\xi - z)^2 + \eta^2} - (\xi - z)},
\end{array} \right.
\end{align}
which is smooth in $t$. Let us also denote $Q_3 = \alpha + i \beta$ and $Q_4 = \alpha - i \beta$ with real-valued
$\alpha$ and $\beta$. The exact solution of the quadrature (\ref{Q-integral}) with
\begin{equation*}
F(Q) = (Q-Q_1) (Q-Q_2) \left[ (Q - \alpha)^2 + \beta^2 \right]
\end{equation*}
for $Q(x,t) \in [Q_2,Q_1]$ is given by
\begin{equation}
Q(x,t) = Q_1 + \frac{(Q_2-Q_1) (1 - {\rm cn}(\nu x; \kappa))}{1 + \gamma + (\gamma - 1) {\rm cn}(\nu x;\kappa)},
\label{Jacobi-2}
\end{equation}
where $\gamma > 0$, $\nu > 0$, and $\kappa \in (0,1)$ are parameters given by
\begin{eqnarray*}
\gamma = \frac{\sqrt{(Q_2-\alpha)^2 + \beta^2}}{\sqrt{(Q_1-\alpha)^2 + \beta^2}}, \\
\end{eqnarray*}
and
\begin{equation*}
\left\{ \begin{array}{l}
\nu^2 = \sqrt{\left[ (Q_1-\alpha)^2 + \beta^2 \right] \left[ (Q_2-\alpha)^2 + \beta^2 \right]}, \\
2 \kappa^2 = 1 - \frac{(Q_1-\alpha) (Q_2 - \alpha) + \beta^2}{\sqrt{\left[ (Q_1-\alpha)^2 + \beta^2 \right] \left[ (Q_2-\alpha)^2 + \beta^2 \right]}}, \end{array} \right. \quad \Rightarrow \quad
\left\{ \begin{array}{l}
\nu^2 = 4 \sqrt{(z_1 - \xi)^2 + \eta^2}, \\
2 \kappa^2 = 1 - \frac{z_1 - \xi}{\sqrt{(z_1 - \xi)^2 + \eta^2}}. \end{array} \right.
\end{equation*}
The function $Q$ is periodic in $x$ with the period $L = 4 K(\kappa)/\nu$
and periodic in $t$ with the period $T = 8 K(k)/\mu$ thanks to the $T$-periodicity of
$z(t)$ and $\dot{\delta}(t)$ in (\ref{Q-roots-complex-correct}).

At $t = 0$ it follows that $\delta(0) = 0$ while $\theta(0) = 0$ can be chosen
without loss of generality.
In this case, the real-valued function $\psi(x,0) = Q(x,0)$ in (\ref{double-periodic})
coincide with the general periodic wave solution of the modified KdV equation
considered in \cite{CPgardner}. The roots of the polynomial $\tilde{P}(\lambda)$ in (\ref{Polynomial-3-explicit})
can then be written in the form:
\begin{equation}
\label{eig-Q-real-complex}
\lambda_1^\pm = \pm\sqrt{z_1} = \pm\frac{Q_1 + Q_2}{2} \biggr|_{t = 0}.
\end{equation}
and
\begin{equation}
\label{eig-Q-complex}
\lambda_2^\pm = \pm\sqrt{\xi + i \eta} = \pm \left[ \frac{1}{4} (Q_1 - Q_2) + \frac{i}{2} \beta \right] \biggr|_{t = 0}, \;\;
\lambda_3^\pm = \pm\sqrt{\xi - i \eta} = \pm \left[ \frac{1}{4} (Q_1 - Q_2) -
\frac{i}{2} \beta \right] \biggr|_{t = 0}.
\end{equation}
For $t \neq 0$, the complex-valued function $\psi(x,t)$ in (\ref{double-periodic}) describes
a double-periodic solution to the NLS equation (\ref{nls}). In the special case $z_1 = 2\xi$ and $\xi^2 + \eta^2 = \frac{1}{4}$,
the exact solution can be simplified to the form (\ref{solA}) derived in \cite{Nail1}. Appendix B
gives technical details of the relevant transformations.

\begin{figure}[h!]
	\centering
	\includegraphics[width=8.5cm,height=6.5cm]{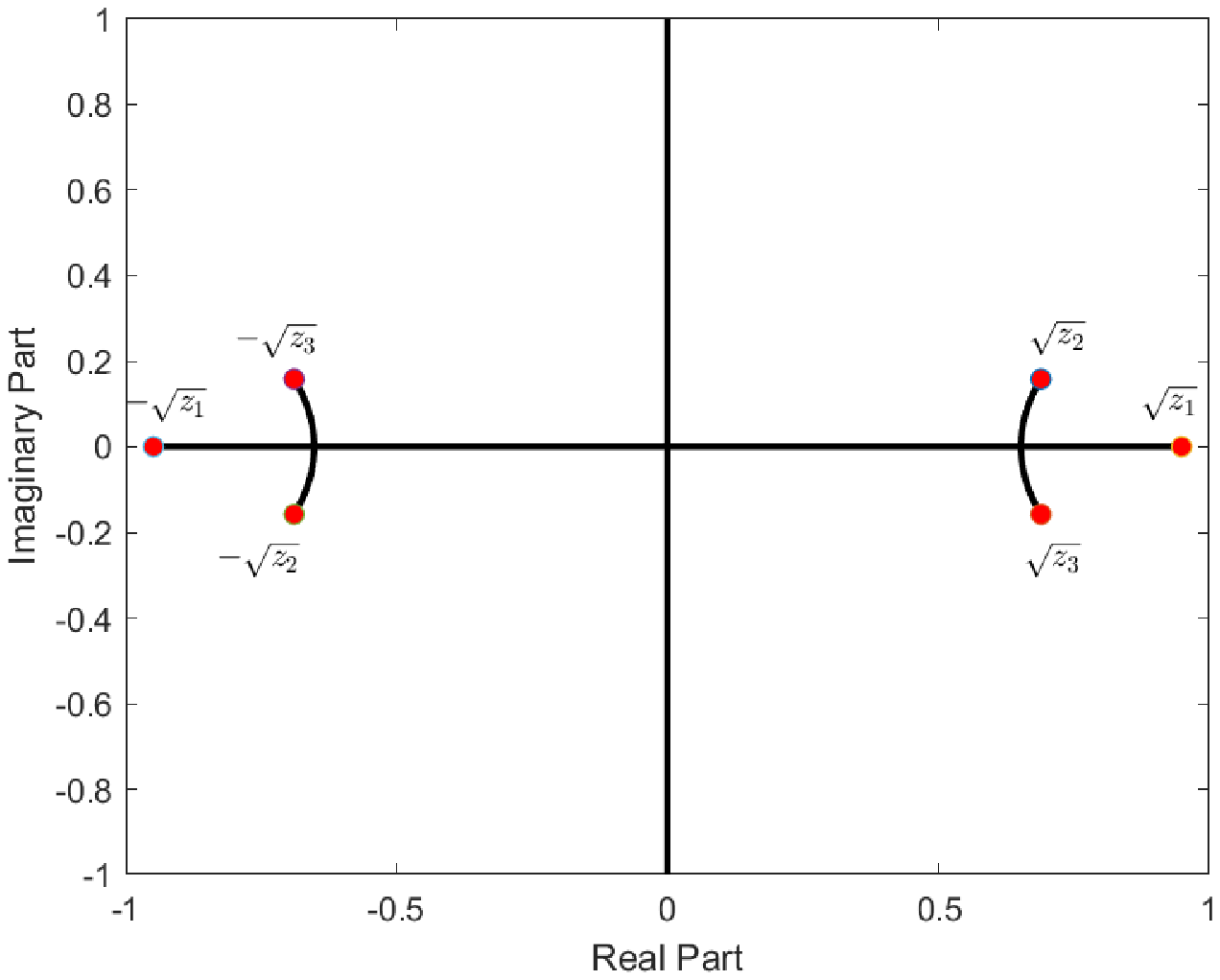}
	\includegraphics[width=8.5cm,height=6.5cm]{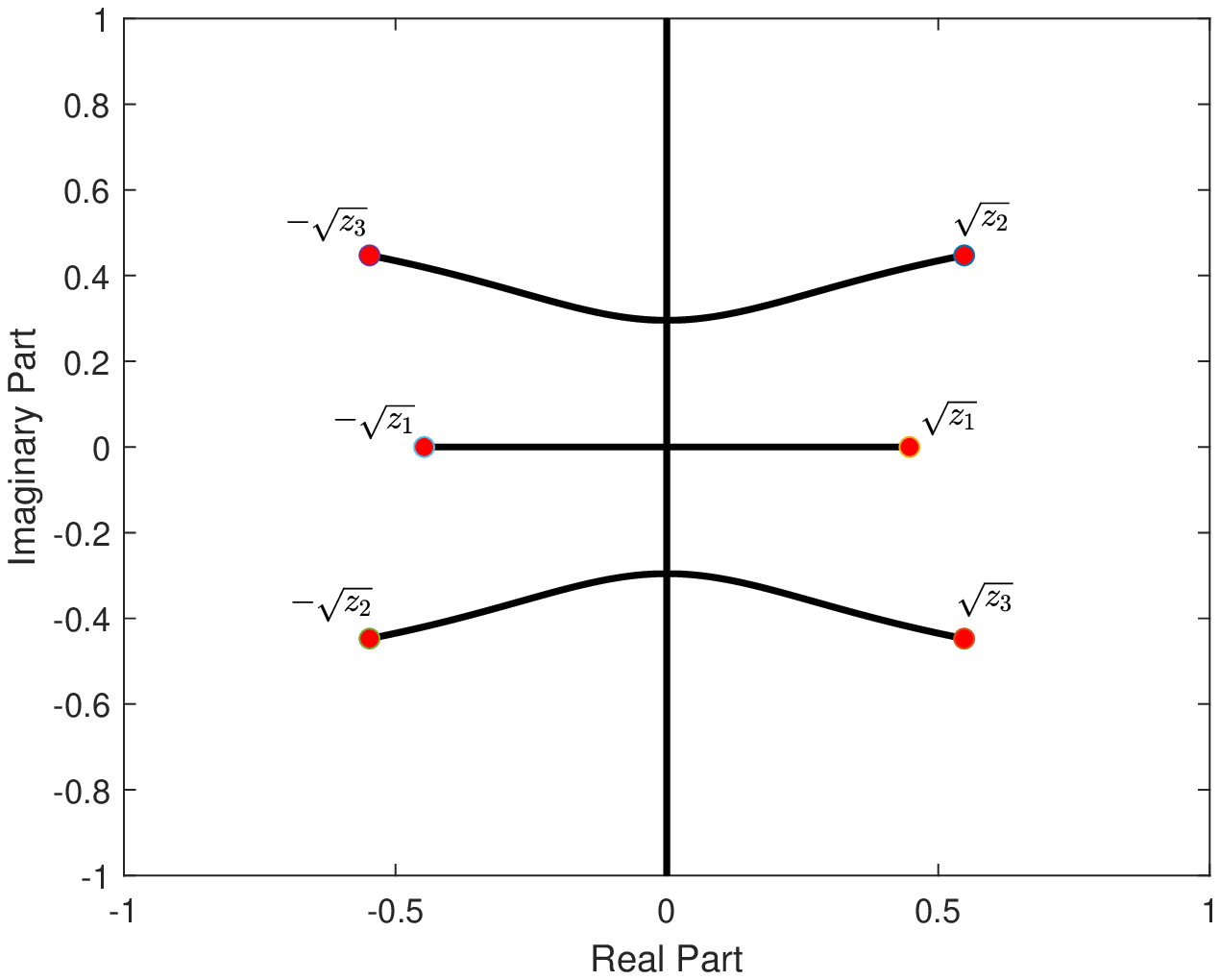}
	\caption{Lax spectrum for the double-periodic solution (\ref{solA}) with
	$k=0.8$ (left) and $k=0.2$ (right). Red dots represent
    eigenvalues $\pm \sqrt{z_1}$, $\pm \sqrt{z_2}$, and $\pm \sqrt{z_3}$..}
	\label{f7}
\end{figure}

Figure \ref{f7} represents the Lax spectrum
computed numerically by using the Floquet--Bloch decomposition
of solutions to the spectral problem (\ref{3.1}) with the periodic potential $u$
in the form (\ref{solA}) for two choices of $k \in (0,1)$. The black curves represent the purely continuous spectrum whereas
the red dots represent eigenvalues (\ref{eig-Q-real-complex}) and (\ref{eig-Q-complex}).
Different reconnections between the bands of the continuous spectrum is observed for $k = 0.8$ (left)
and $k = 0.2$ (right).

\section{Rogue waves on the double-periodic background}
\label{sec-eigenvectors}

Here we characterize the squared eigenfunctions of the Lax operators in
(\ref{3.1})--(\ref{3.2})
in terms of the solutions $u$ to the third-order Lax--Novikov equation (\ref{LN-3}).
For each pair of admissible eigenvalues $\lambda_1$ and $\lambda_2$ among roots
of the polynomial $\tilde{P}(\lambda)$ in (\ref{Polynomial-3-explicit}), the squared eigenfunctions
are double-periodic functions with the same periods as the solution $u$. The second, linearly independent solution
of the linear equations (\ref{3.1})--(\ref{3.2}) exist for the same eigenvalues
and we characterize the second solution in terms of the double-periodic eigenfunctions
similarly to our previous work \cite{CPWnls}. The second solution is generally non-periodic
but linearly growing in variables $(x,t)$.

Let us recall the representations (\ref{W-11-12}), (\ref{3.19}), and (\ref{3.17})
for $W_{11}(\lambda)$ and $W_{12}(\lambda)$ in terms of the squared eigenfunctions
and the periodic solution $u$. Also recall that $S_3 = T_4 = 0$ for the third-order
Lax--Novikov equation (\ref{LN-3}), which we take with $a = c = e = 0$.
By computing and comparing the residues of expressions (\ref{W-11-12}), (\ref{3.19}), and (\ref{3.17}) at the simple poles
$\lambda = \lambda_1$ and $\lambda = -\bar{\lambda}_1$, we obtain the explicit expressions:
\begin{align}
\label{eig-4}
\left\{ \begin{array}{l}
p_1^2 = \frac{\lambda_1}{4 (\lambda_1 + \bar{\lambda}_1) (\lambda_1 - \lambda_2) (\lambda_1 + \bar{\lambda}_2)}
\left[ u'' + 2 |u|^2 u + 4 (b + \lambda_1^2) u + 2 \lambda_1 u' \right], \\
q_1^2 = \frac{\lambda_1}{4 (\lambda_1 + \bar{\lambda}_1) (\lambda_1 - \lambda_2) (\lambda_1 + \bar{\lambda}_2)}
\left[ \bar{u}'' + 2 |u|^2 \bar{u} + 4 (b + \lambda_1^2) \bar{u}
- 2 \lambda_1 \bar{u}' \right], \\
p_1 q_1 = -\frac{\lambda_1}{4 (\lambda_1 + \bar{\lambda}_1) (\lambda_1 - \lambda_2) (\lambda_1 + \bar{\lambda}_2)}
\left[ u' \bar{u} - u \bar{u}' + 2 \lambda_1 (2 b + 2 \lambda_1^2 + |u|^2)
\right]. \end{array} \right.
\end{align}
Expressions for $p_2^2$, $q_2^2$, and $p_2 q_2$ are obtained from (\ref{eig-4}) after replacing
$\lambda_1$ and $\lambda_2$. There exist three admissible pairs of eigenvalues given by roots $\pm \sqrt{z_1}$,
$\pm \sqrt{z_2}$, and $\pm \sqrt{z_3}$ of the polynomial $\tilde{P}(\lambda)$ in (\ref{Polynomial-3-explicit})
and each pair of roots can be taken in place of $\lambda_1$ and $\lambda_2$.

\begin{figure}[h!]
	\centering
	\includegraphics[width=8.5cm,height=6.5cm]{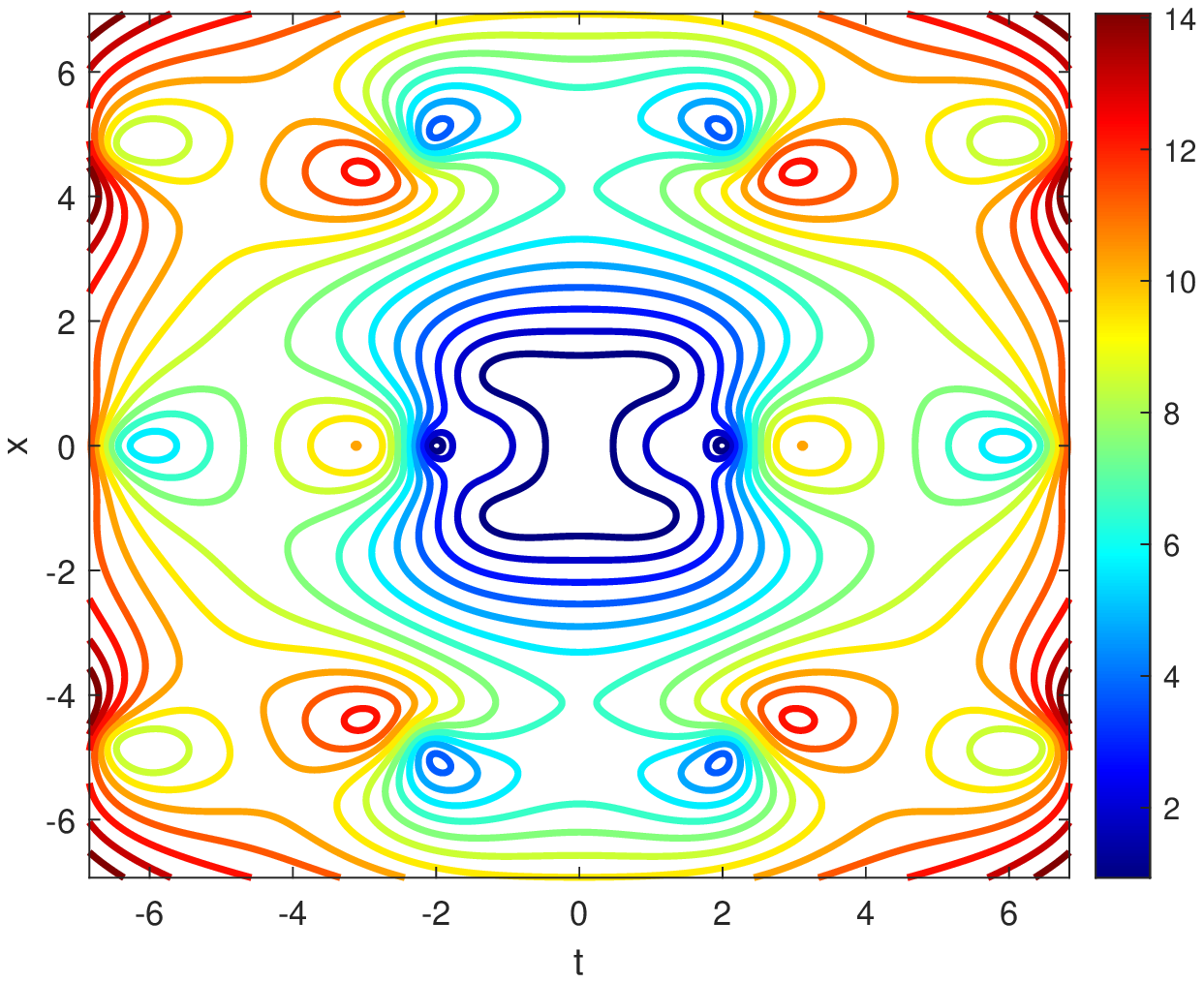}
	\includegraphics[width=8.5cm,height=6.5cm]{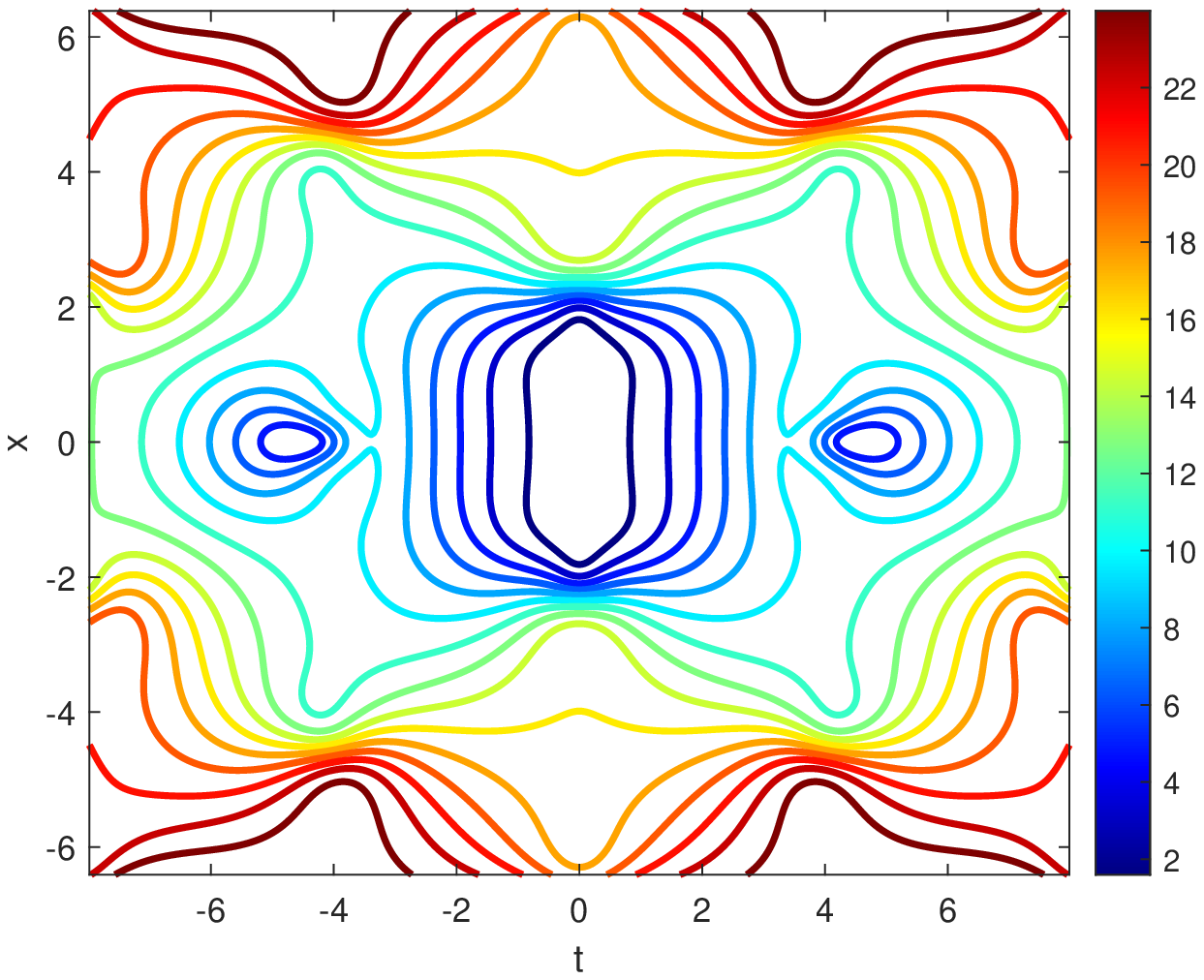}
	\caption{The level set of $|\phi_{1}|$ generated for the double-periodic solution (\ref{solB})
		with $k=0.9$ (left) and the double-periodic solution (\ref{solA}) with $k = 0.8$ (right) for the first eigenvalue $\sqrt{z_1}$. }
	\label{f10}
\end{figure}

Let $\varphi = (p_1,q_1)^T$ be a solution of the linear equations (\ref{3.1})--(\ref{3.2}) for $\lambda = \lambda_1$.
The second, linearly independent solution $\varphi = (\hat{p}_1,\hat{q}_1)^T$ of the same equations
is obtained in the form:
\begin{equation}
\label{represent-new}
\hat{p}_1 = p_1 \phi_1 - \frac{2 \bar{q}_1}{|p_1|^2 + |q_1|^2}, \quad
\hat{q}_1 = q_1 \phi_1 + \frac{2 \bar{p}_1}{|p_1|^2 + |q_1|^2},
\end{equation}
where $\phi_1$ is to be determined. Wronskian between the two solutions is normalized by
$p_1 \hat{q}_1 - \hat{p}_1 q_1 = 2$. Substituting (\ref{represent-new}) into (\ref{3.1})
and using (\ref{3.1}) for $\varphi = (p_1,q_1)^T$ yield the following first-order equation for $\phi_1$:
\begin{equation}
\label{phi-der-x}
\frac{\partial \phi_1}{\partial x} = F := -\frac{4 (\lambda_1 +
\bar{\lambda}_1) \bar{p}_1 \bar{q}_1}{(|p_1|^2+|q_1|^2)^2}.
\end{equation}
Similarly, substituting (\ref{represent-new}) into (\ref{3.2})
and using (\ref{3.2}) for $\varphi = (p_1,q_1)^T$ yield another equation for $\phi_1$:
\begin{equation}
\label{phi-der-t}
\frac{\partial \phi_1}{\partial t} = G := -\frac{4 i (\lambda_1^2 - \bar{\lambda}_1^2) \bar{p}_1 \bar{q}_1}{(|p_1|^2+|q_1|^2)^2}
+ \frac{2i (\lambda_1 + \bar{\lambda}_1) (u \bar{p}_1^2 + \bar{u}
\bar{q}_1^2)}{(|p_1|^2+|q_1|^2)^2}.
\end{equation}
The system of first-order equations (\ref{phi-der-x}) and (\ref{phi-der-t}) is compatible in the sense
$F_t = G_x$ since it is derived from the compatible Lax system (\ref{3.1})--(\ref{3.2}).
Therefore, it can be solved with the explicit integration formula:
\begin{equation}
\label{phi}
\phi_1(x,t) = \int_{x_0}^x F(x',t) dx' + \int_{t_0}^t G(x_0,t') dt',
\end{equation}
where $(x_0,t_0)$ is arbitrarily fixed. Figure \ref{f10} shows contour plots of $|\phi_{1}|$ on the $(x,t)$ plane
generated for the double-periodic solutions with the same parameters as on Figure \ref{f1}. The choice $\lambda_1 = \sqrt{z_1}$
is used here. The value of $|\phi_1|$
increases as $(x,t)$ deviate further away from ($x_{0},t_{0}$), which is placed at the origin. We have checked that the same behavior
holds for every choice of $k \in (0,1)$ in the double-periodic solutions (\ref{solB}) and (\ref{solA})
and for other two eigenvalues $\sqrt{z_2}$ and $\sqrt{z_3}$.

\begin{figure}[!htb]
	\centering
	\includegraphics[width=8cm,height=6cm]{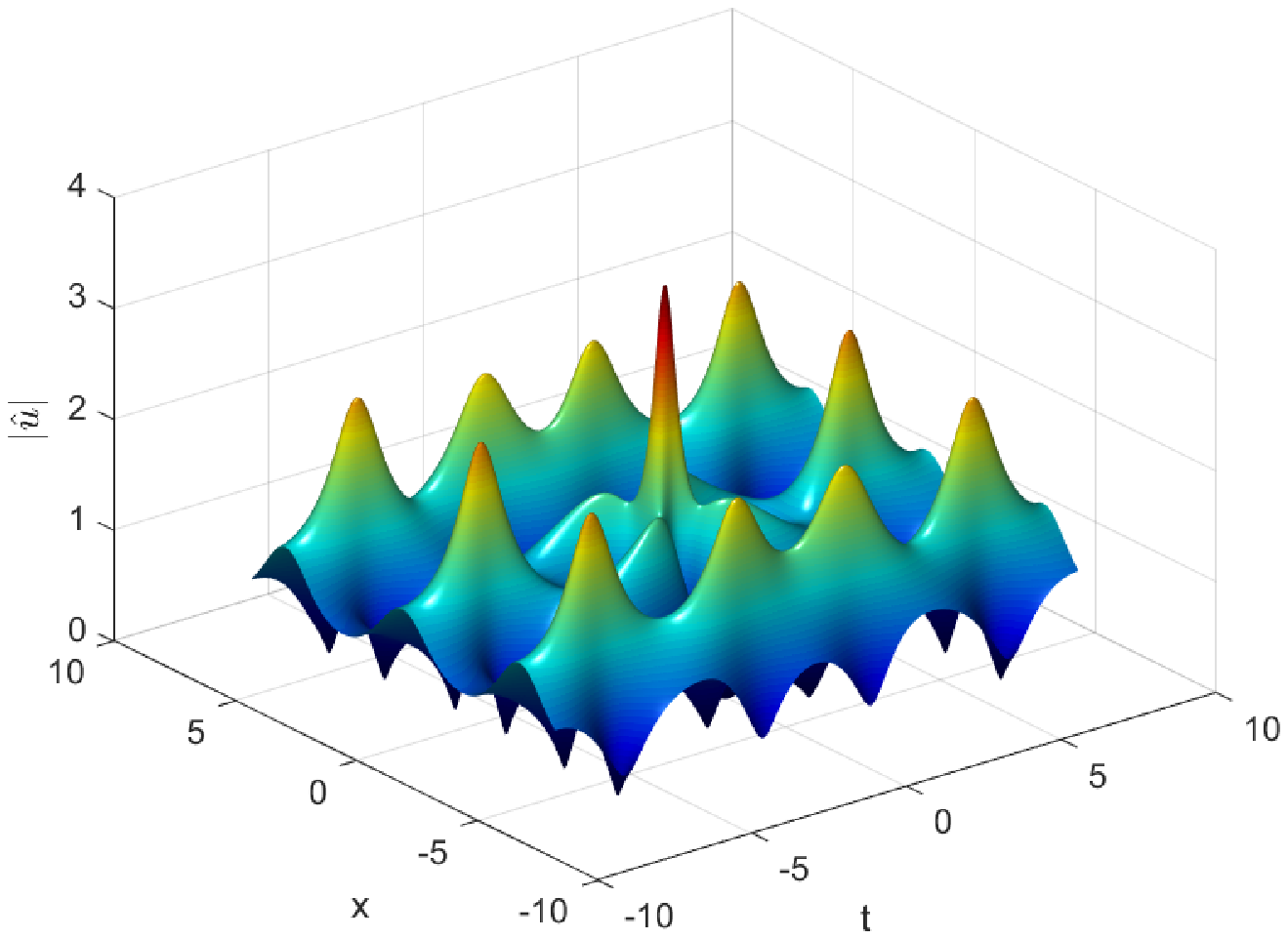}
	\includegraphics[width=8cm,height=6cm]{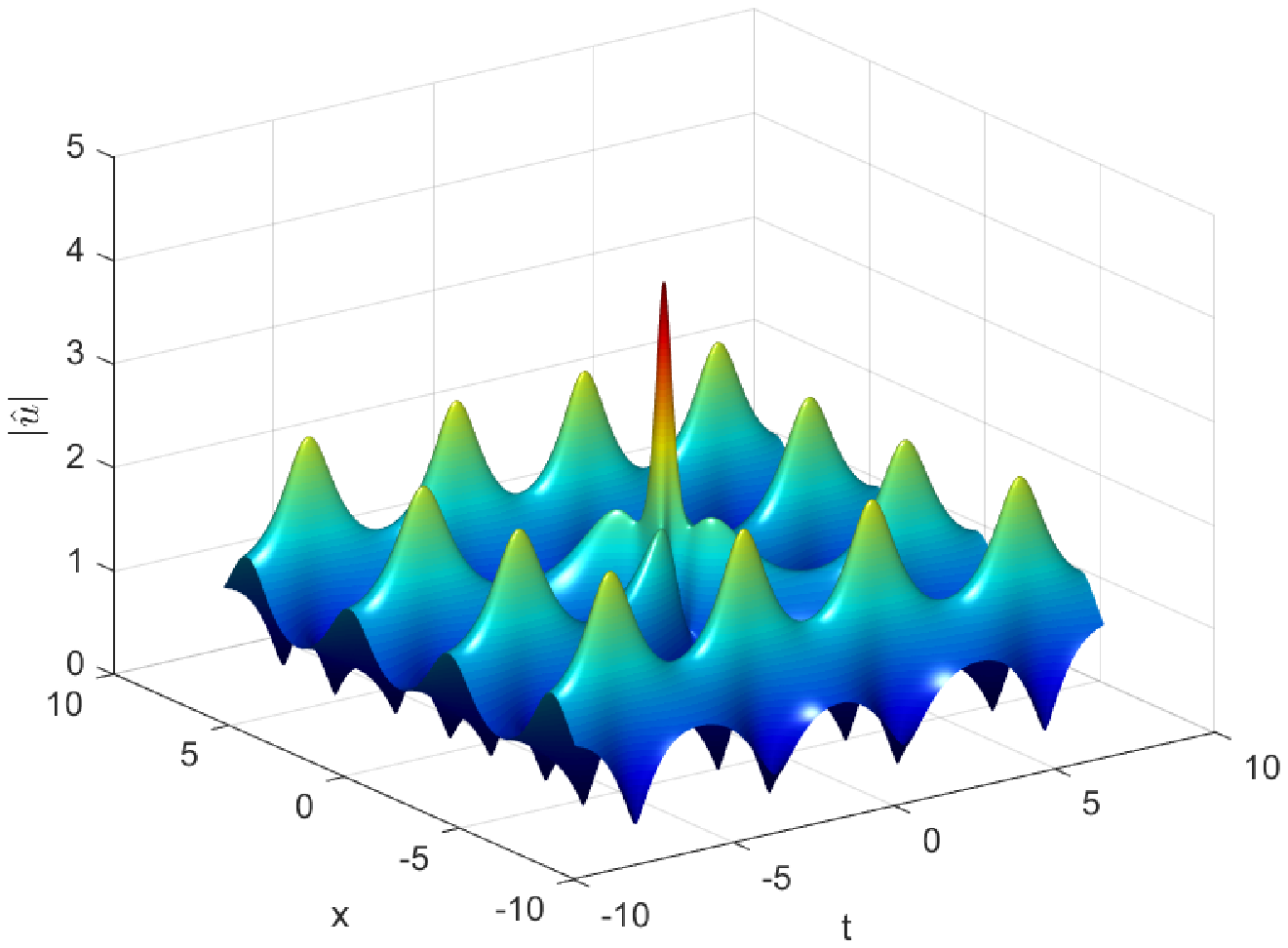}\\
	\includegraphics[width=8cm,height=6cm]{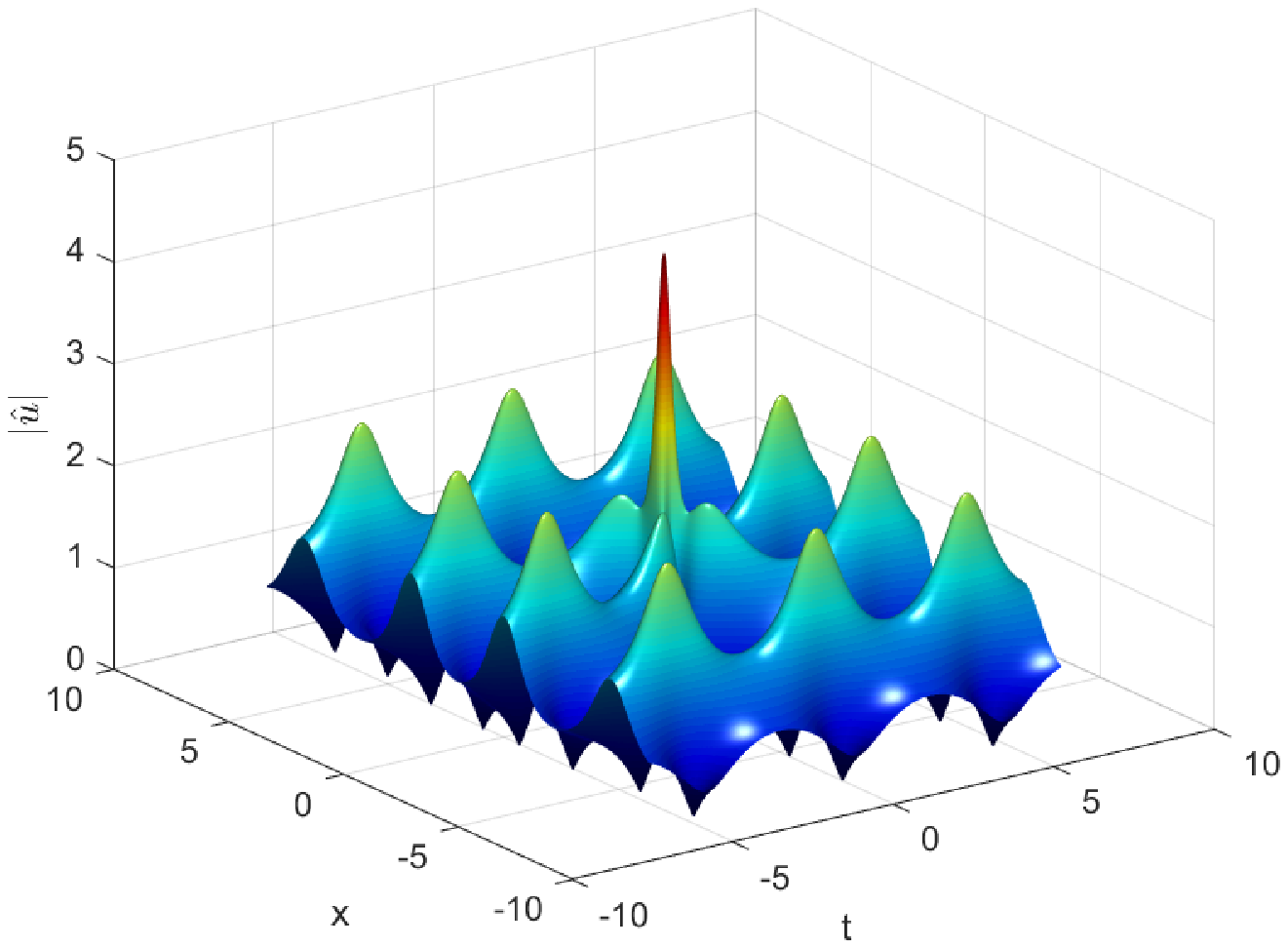}
	\caption{Rogue waves on the background of the double-periodic solution
	(\ref{solB}) with $k=0.9$ for eigenvalues $\lambda_{1} = \sqrt{z_{1}}$
		(top left), $\lambda_{2} = \sqrt{ z_{2} }$ (top right),
        and $\lambda_{3} = \sqrt{z_{3}}$ (bottom). In all cases, we set
        $(x_0,t_0) = (0,0)$.}
	\label{12rogue}
\end{figure}

By using the one-fold Darboux transformation (\ref{1-fold}) with the second solution
$\varphi = (\hat{p}_1,\hat{q}_1)^t$ of the linear equations (\ref{3.1})--(\ref{3.2}) with $\lambda = \lambda_1$,
we obtain a new solution to the NLS equation (\ref{nls}) in the form:
\begin{equation}
\label{1-fold-rogue}
\hat{u} = u + \frac{2 (\lambda_1 + \bar{\lambda}_1) \hat{p}_1
\bar{\hat{q}}_1}{|\hat{p}_1|^2 + |\hat{q}_1|^2}
\end{equation}
This new solution can be rewritten in the explicit form:
\begin{equation}
\label{rogue-4}
\hat{u} = u + \frac{2(\lambda_1 + \bar{\lambda}_1) \left[ p_1 (|p_1|^2 + |q_1|^2) \phi_1 - 2 \bar{q}_1 \right]
\left[ \bar{q}_1 (|p_1|^2 + |q_1|^2) \bar{\phi}_1 + 2 p_1 \right]}{\left| p_1 (|p_1|^2 + |q_1|^2) \phi_1 - 2 \bar{q}_1 \right|^2
+ \left| \bar{q}_1 (|p_1|^2 + |q_1|^2) \bar{\phi}_1 + 2 p_1 \right|^2},
\end{equation}
where $p_1^2$, $q_1^2$, and $p_1 q_1$ are defined by (\ref{eig-4}). Figure
\ref{12rogue} shows surface plots of $|\hat{u}|$ generated on the
double-periodic background (\ref{solB}) with $k = 0.9$ for three different eigenvalues $\sqrt{z_1}$, $\sqrt{z_2}$, and $\sqrt{z_3}$
shown on Figure \ref{f6}.

Notice from Figure \ref{f10} that $|\phi_1(x,t)| \to \infty$ as $|x| + |t| \to \infty$ everywhere on the $(x,t)$-plane.
The explicit representation (\ref{rogue-4}) implies that
\begin{equation}
\label{asymptotic-inf}
\hat{u} |_{|\phi_1| \to \infty} = u + \frac{2 (\lambda_1 + \bar{\lambda}_1) p_1
\bar{q}_1}{|p_1|^2 + |q_1|^2} =: \tilde{u}.
\end{equation}
We have verified numerically that $\tilde{u}$ is a translated version of the original double-periodic solution $u$
due to the symmetries of the NLS equation (\ref{nls}):
\begin{equation}
\label{translations}
|\tilde{u}(x,t)| = |u(x-\beta,t-\tau)|, \quad \beta, \tau \in \mathbb{R}.
\end{equation}
In particular, if $L$ and $T$ are fundamental periods of the double-periodic solution (\ref{solB}) in $x$ and $t$ respectively,
then we have found that $\beta$ and $\tau$ are independent of $k \in (0,1)$. Table \ref{Table0} lists values of $\beta$ and $\tau$
for the three eigenvalues.
\begin{table}[ht]
\begin{center}
\begin{tabular}{|c|c|c|c|}
\hline
Rogue wave & $\beta$ & $\tau$  \\
\hline
$\lambda_1 = \sqrt{z_1}$ & $0$ & $T/2$  \\
$\lambda_2 = \sqrt{z_2}$ & $L/2$ & $T/2$ \\
$\lambda_3 = \sqrt{z_3}$ & $L/2$ & $0$  \\
\hline
\end{tabular}
\end{center}
\caption{Parameters of the translations (\ref{translations}) for the double-periodic wave (\ref{solB}).}
\label{Table0}
\end{table}

\begin{figure}[h!]
	\centering
	\includegraphics[width=8cm,height=5.8cm]{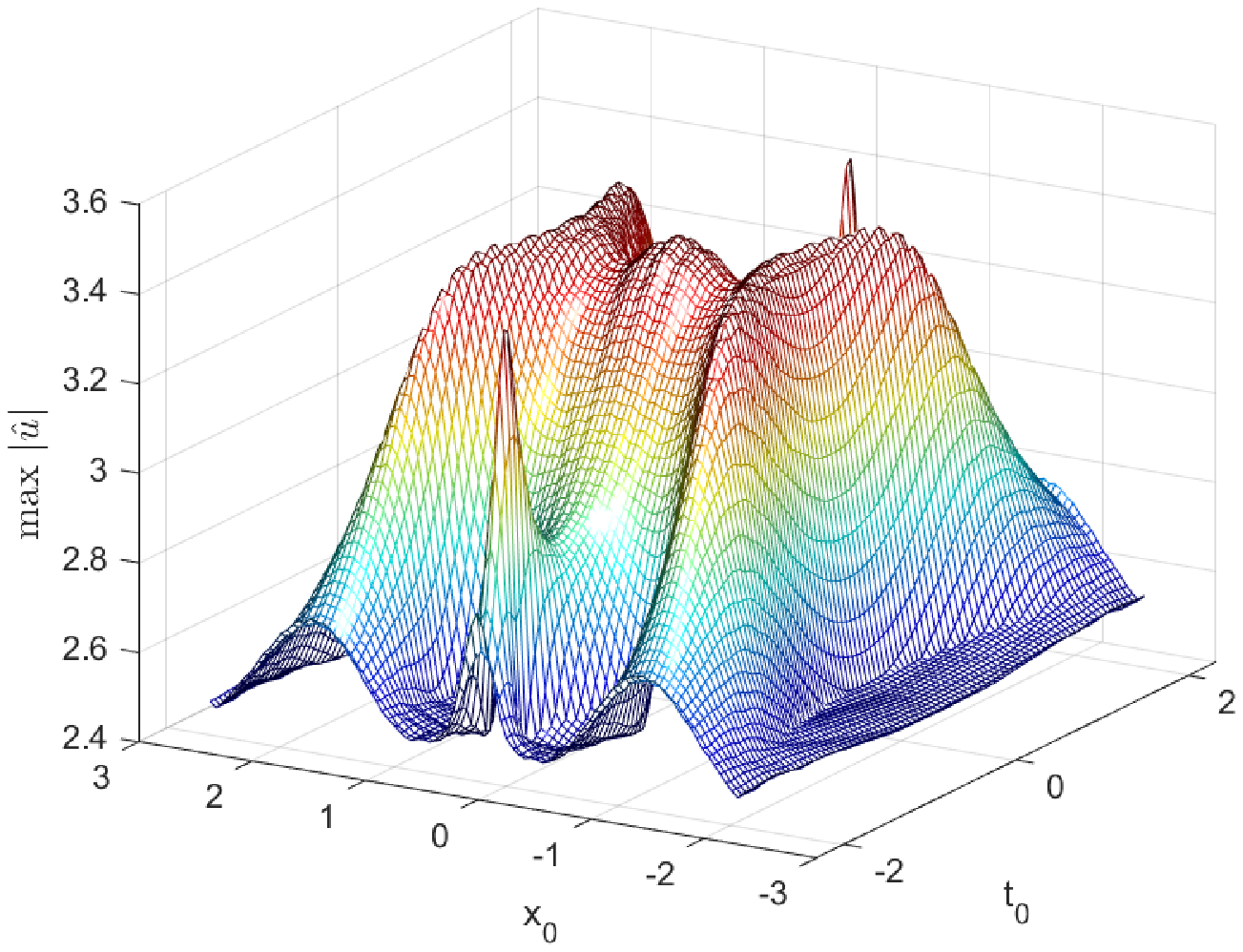}
	\includegraphics[width=8cm,height=5.8cm]{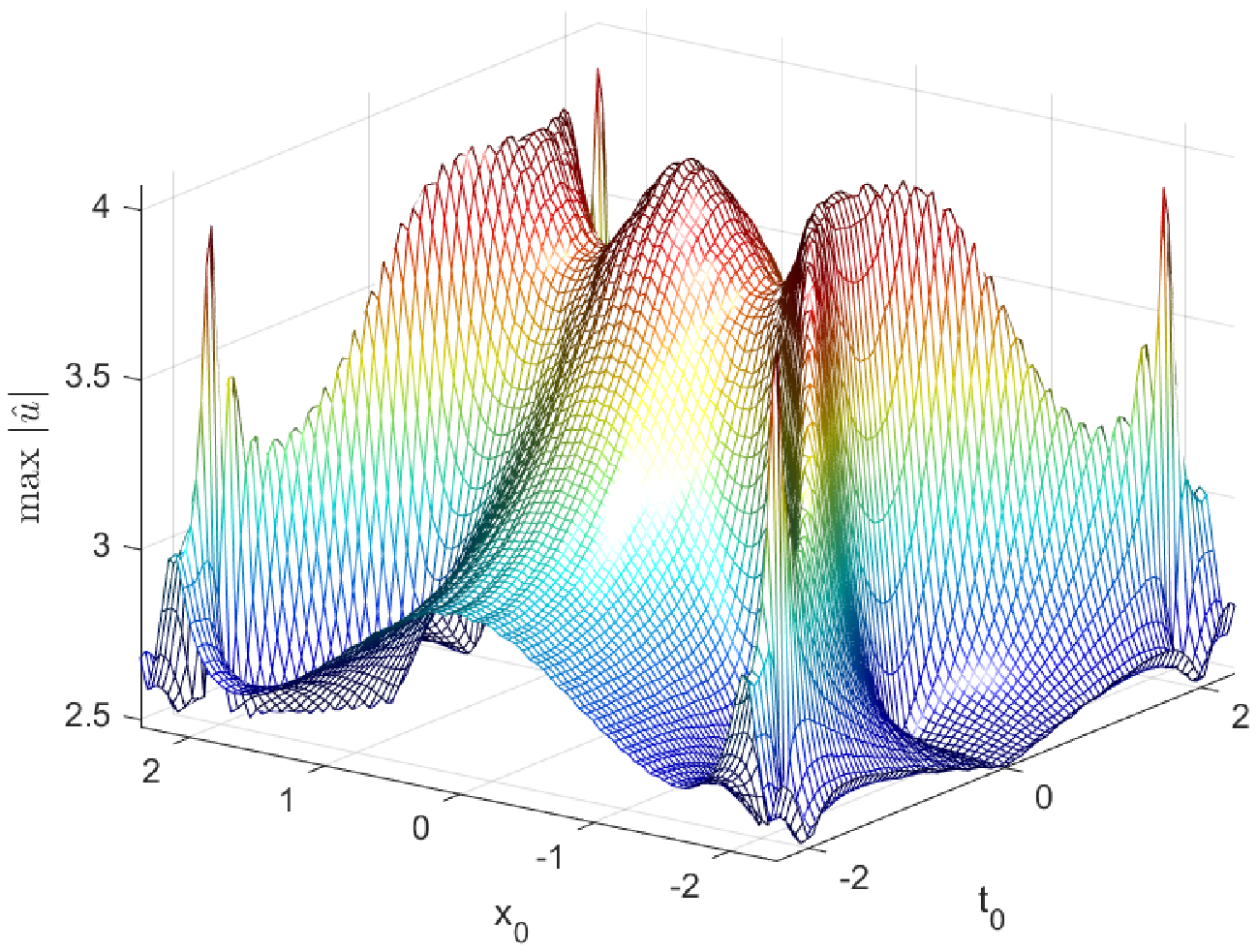}\\
	\includegraphics[width=8cm,height=5.8cm]{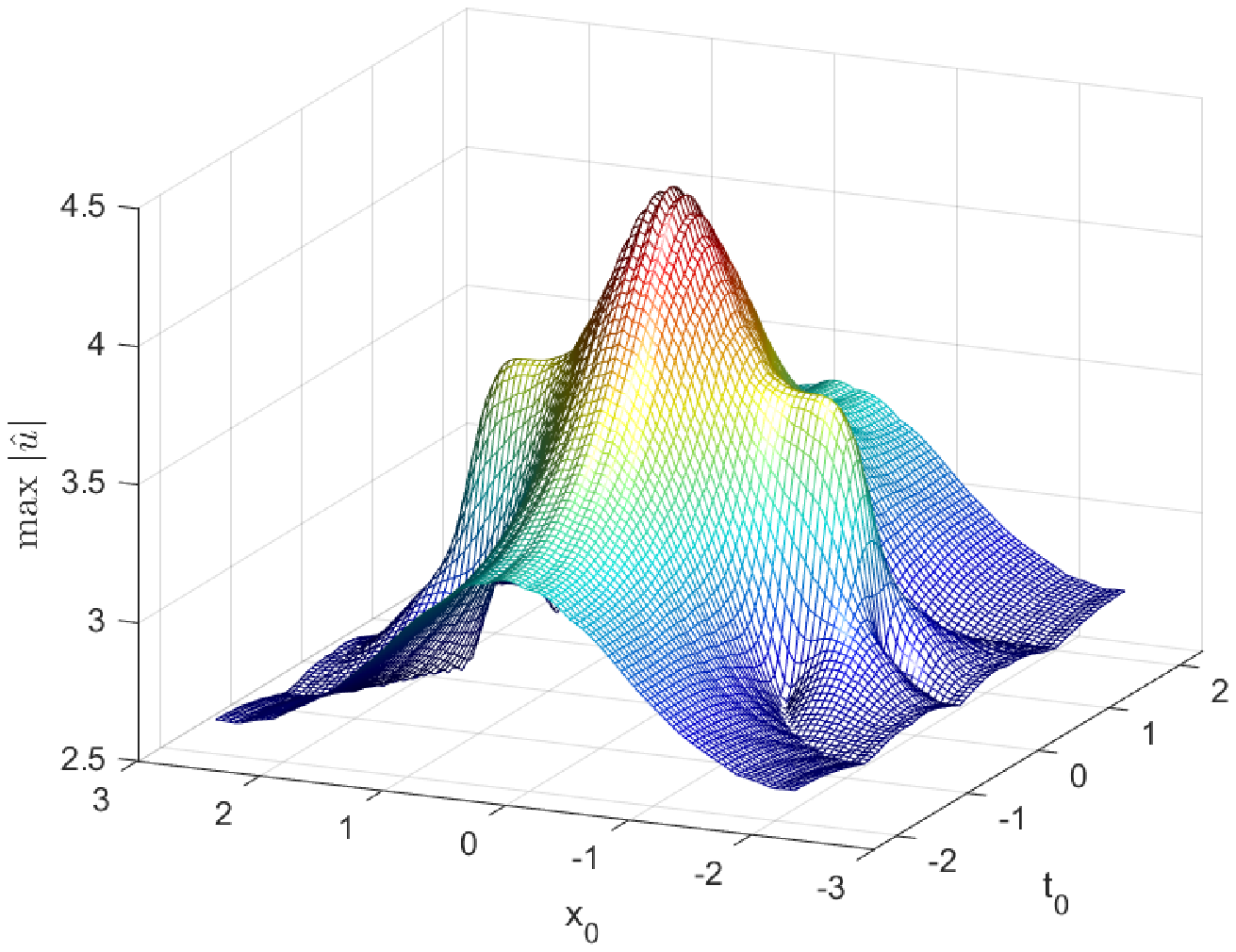}
	\caption{Magnification given by $\max_{(x,t) \in \mathbb{R}^2}
	|\hat{u}(x,t)|$ versus $(x_0,t_0)$
		for rogue waves generated from the double-periodic solution
		(\ref{solB})
		with $k=0.9$ for $\lambda_{1} = \sqrt{ z_{1} }$ (top left),
		$\lambda_{2} = \sqrt{ z_{2} }$ (top right), and $\lambda_{3} = \sqrt{
		z_{3}
		}$ (bottom). }
	\label{fm1}
\end{figure}

In order to study magnification of the rogue wave, we compute $\hat{u}(x_0,t_0)$ for which $\phi_1(x_0,t_0) = 0$,
\begin{equation}
\label{asymptotic-zero}
\hat{u} |_{\phi_1 = 0} = u - \frac{2 (\lambda_1 + \bar{\lambda}_1) p_1 \bar{q}_1}{|p_1|^2 + |q_1|^2}
= 2 u - \tilde{u}.
\end{equation}
Figure \ref{fm1} shows how $\max_{(x,t) \in \mathbb{R}^2} |\hat{u}(x,t)|$ depends on $(x_{0},t_{0})$ for the three rogue waves
on Figure \ref{12rogue}. For $\lambda_3 = \sqrt{z_3}$ (bottom panel), the maximal magnification of $|\hat{u}|$ is reached at $(x_0,t_0) = (0,0)$,
for which the argument of $\max_{(x,t) \in \mathbb{R}^2} |\hat{u}(x,t)|$ occurs at $(x,t) = (0,0)$, where
the exact formula (\ref{asymptotic-zero}) can be used. For $\lambda_1 = \sqrt{z_1}$ and $\lambda_2 = \sqrt{z_2}$
(top panel), the maximal magnification occurs at $(x_0,t_0) = (0,0)$ and
additional points , for which $\max_{(x,t) \in \mathbb{R}^2} |\hat{u}(x,t)|$ is
nearly the same up to numerical errors.
We have checked that the rogue waves with $(x_0,t_0) \neq (0,0)$ look similar
to those on Figure \ref{12rogue} with $(x_0,t_0) = (0,0)$
and the maximal value is still attained at the central peak near $(x,t) =
(0,0)$.

\begin{figure}[h!]
	\centering
	\includegraphics[width=8cm,height=6cm]{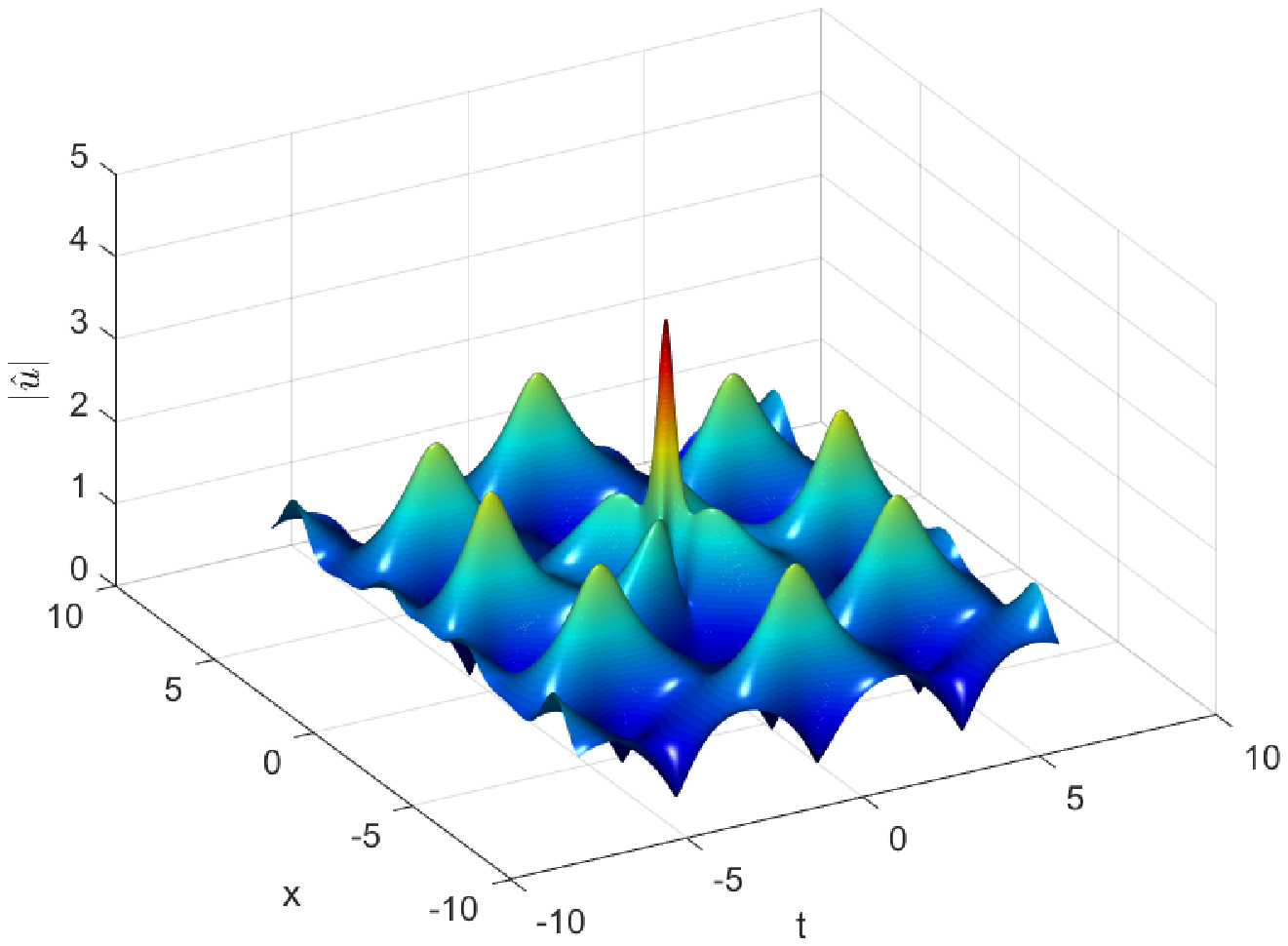}
	\includegraphics[width=8cm,height=6cm]{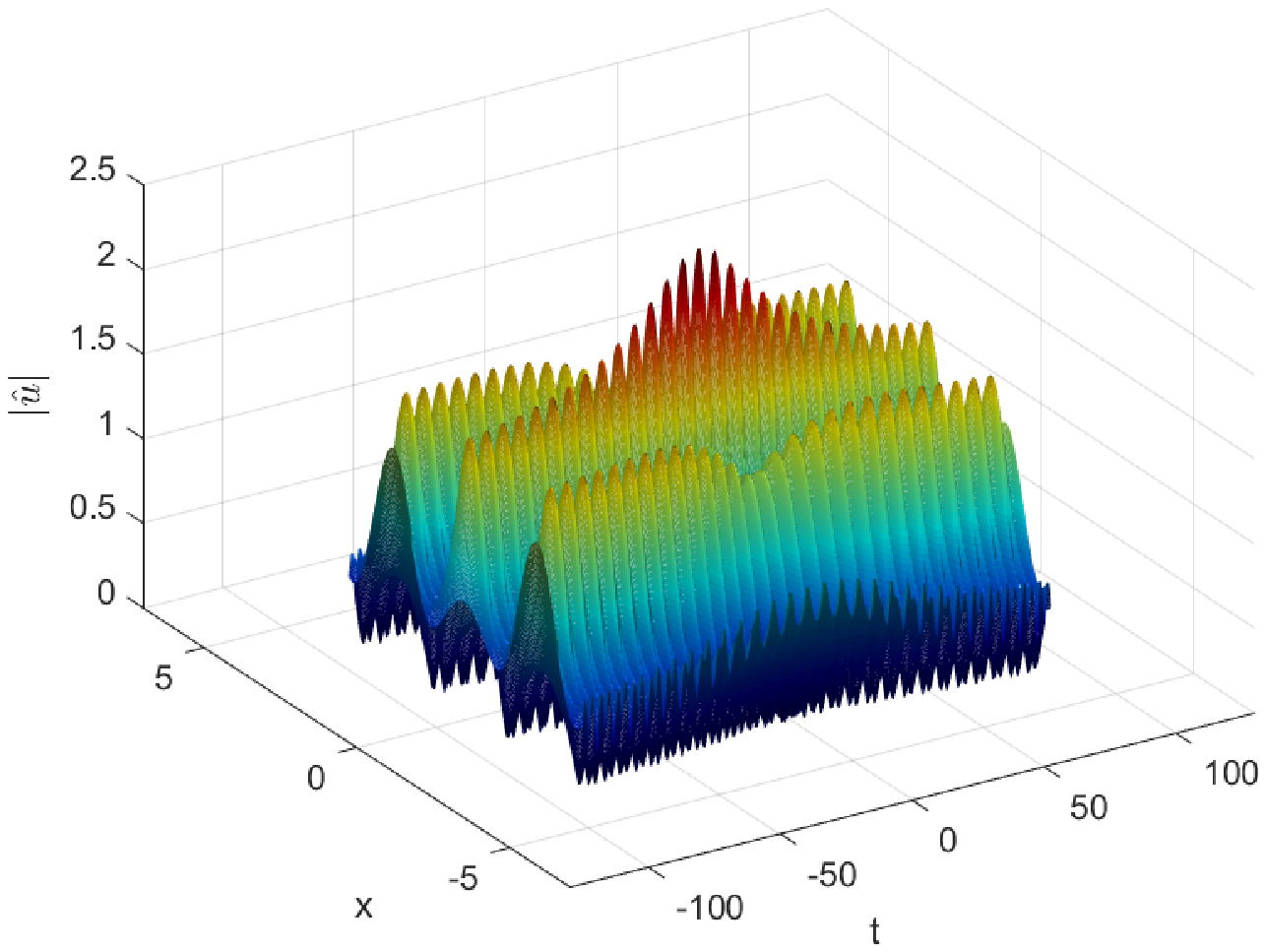}
	\includegraphics[width=8cm,height=6cm]{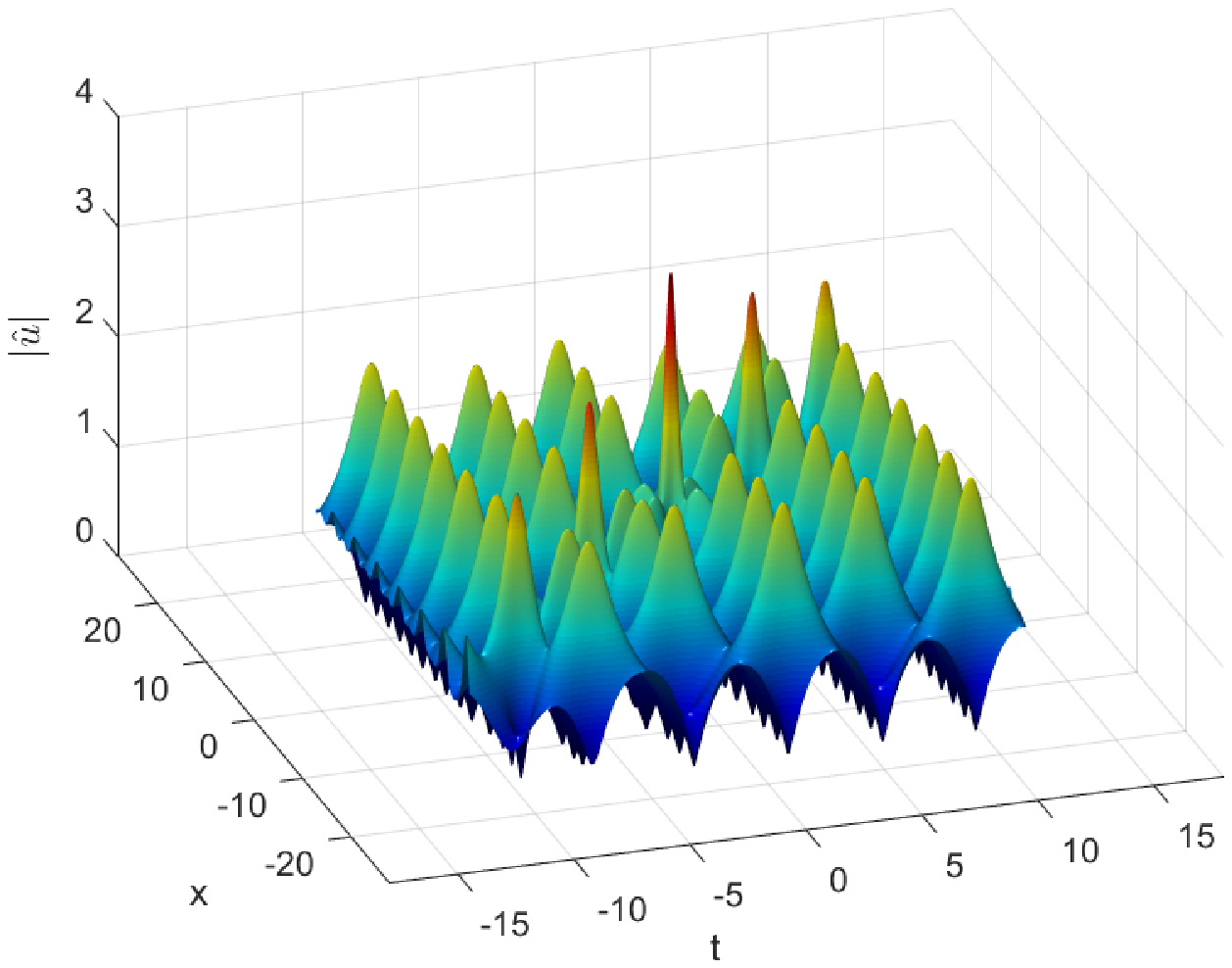}
	\includegraphics[width=8cm,height=6cm]{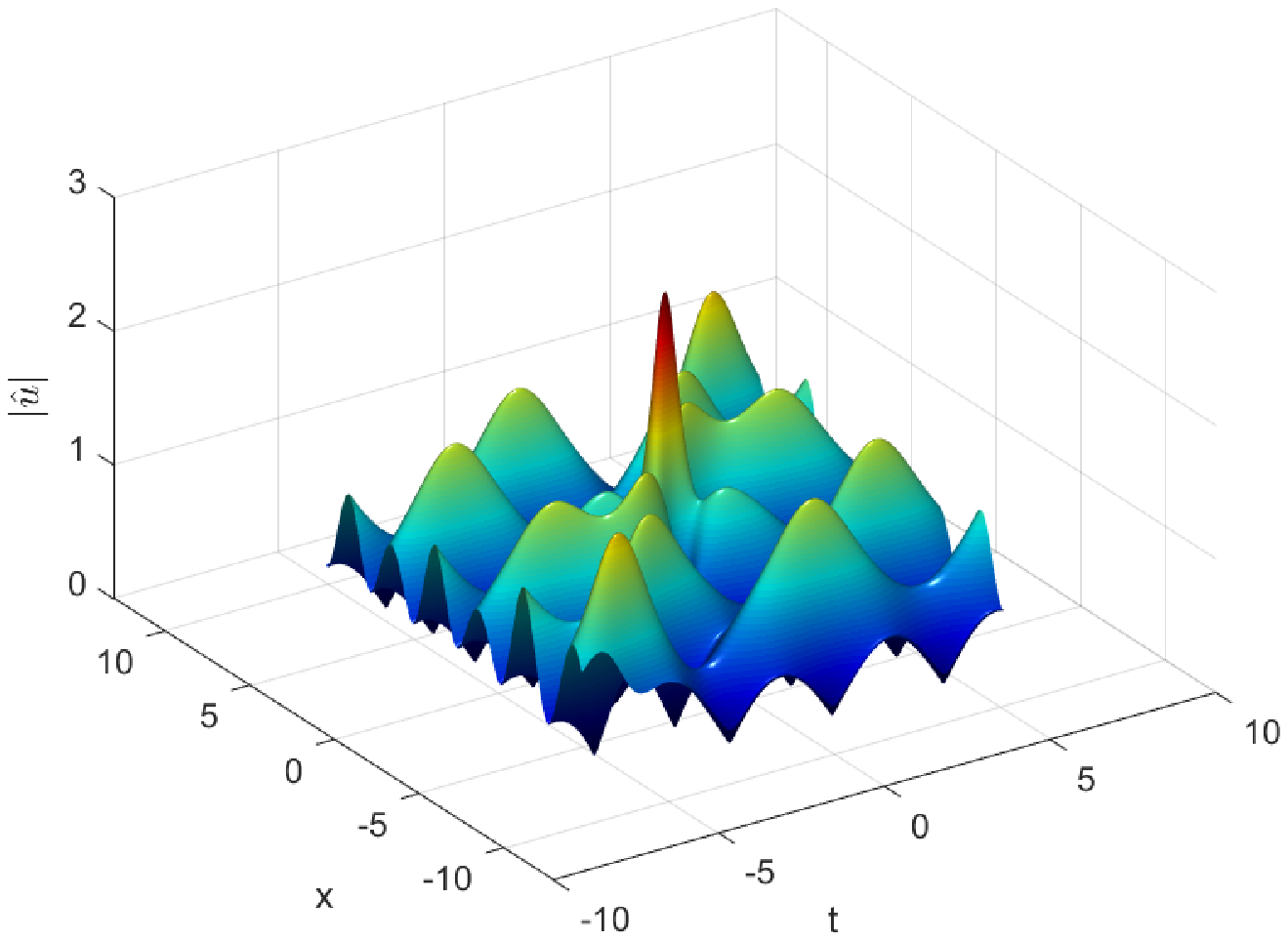}
	\caption{Rogue waves generated on the background of the double-periodic solution (\ref{solA})
		with $k=0.8$ (left) and $k=0.2$ (right) for the real eigenvalue $\lambda_{1} =
		\sqrt{ z_{1} }$ (top) and the complex eigenvalue $\lambda_{2} = \sqrt{
		z_{2} }$ (bottom). In all cases, we set $(x_0,t_0) = (0,0)$. }
	\label{f9}
\end{figure}

Let us now reproduce similar results for the double-periodic background (\ref{solA}).
Figure \ref{f9} shows the surface plots of $|\hat{u}|$ for the rogue waves
generated for the double-periodic solution (\ref{solA}) with $k = 0.8$ (left) and $k = 0.2$ (right)
for real eigenvalue $\lambda_1 = \sqrt{z_1}$ (top) and complex eigenvalue $\lambda_2 = \sqrt{\xi + i \eta}$ (bottom)
shown on Figure \ref{f7}. Rogue waves for the complex-conjugate eigenvalue $\lambda_3 = \sqrt{\xi - i \eta}$
are similar to those for $\lambda_2 = \sqrt{\xi + i \eta}$ thanks to symmetries of the NLS equation (\ref{nls}).
Note that the maximum of $|\hat{u}|$ is bigger for $\lambda_1$ if $k = 0.8$ and for $\lambda_2$ if $k = 0.2$.
Comparison with Fig. \ref{f7} indicates that the rogue wave is larger if it is associated with the eigenvalue,
which is most distant from the imaginary axis ${\rm Re}(\lambda) = 0$

We have checked again that $\tilde{u}$ in (\ref{asymptotic-inf}) is the translation
of the original double-periodic wave $u$ given by (\ref{translations}) with parameters
$\beta$ and $\tau$ given in Table \ref{Table0b} for every $k \in (0,1)$.
\begin{table}[ht]
\begin{center}
\begin{tabular}{|c|c|c|c|}
\hline
Rogue wave & $\beta$ & $\tau$ \\
\hline
$\lambda_1 = \sqrt{z_1}$ & $L/2$ & $0$  \\
$\lambda_2 = \sqrt{\xi + i \eta}$ & $L/4$ & $3T/4$  \\
\hline
\end{tabular}
\end{center}
\caption{Parameters of the translations (\ref{translations}) for the double-periodic wave (\ref{solA}).}
\label{Table0b}
\end{table}

\begin{figure}[h!]
	\centering
	\includegraphics[width=7.5cm,height=5.4cm]{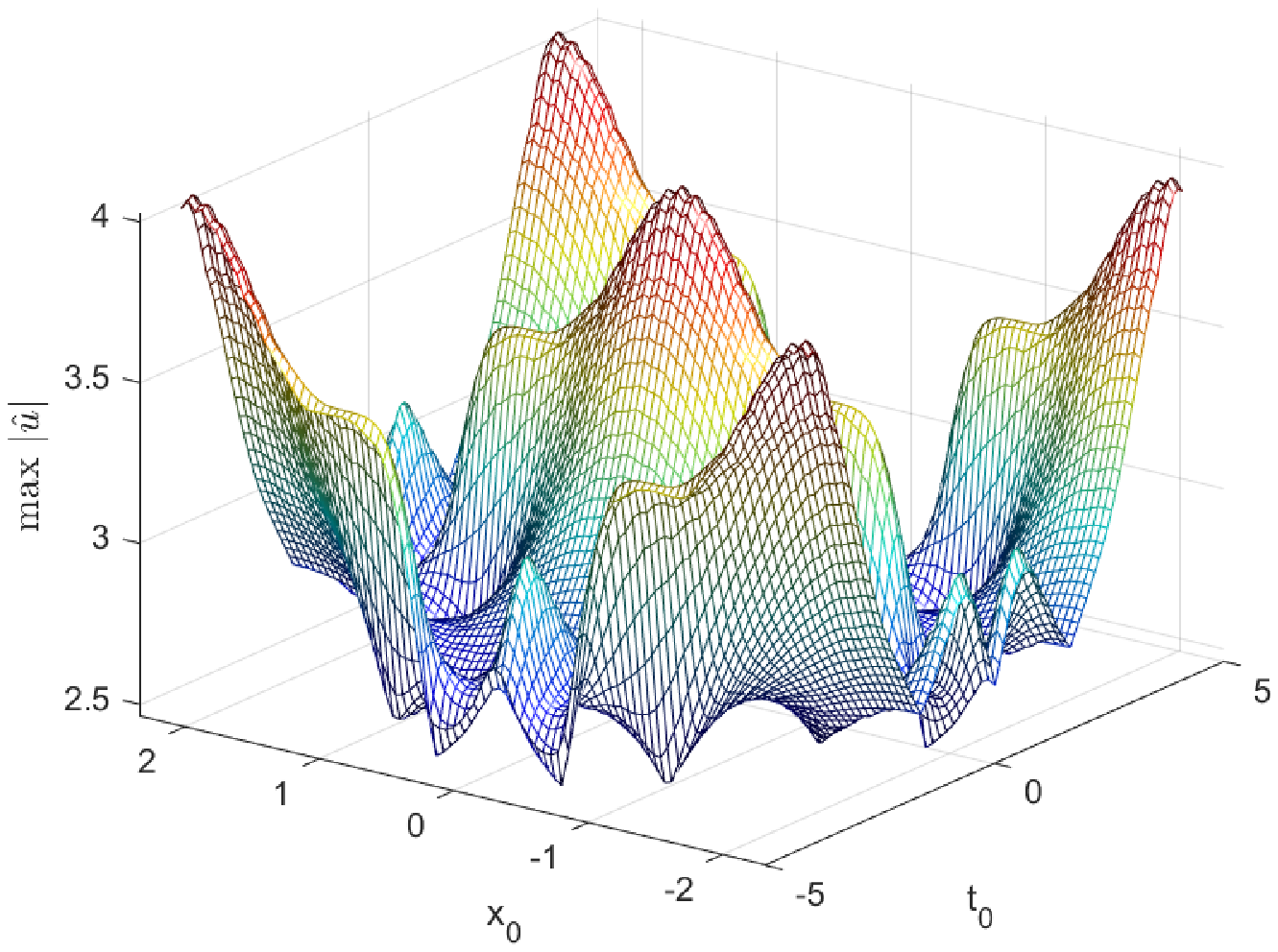}
	\includegraphics[width=7.5cm,height=5.4cm]{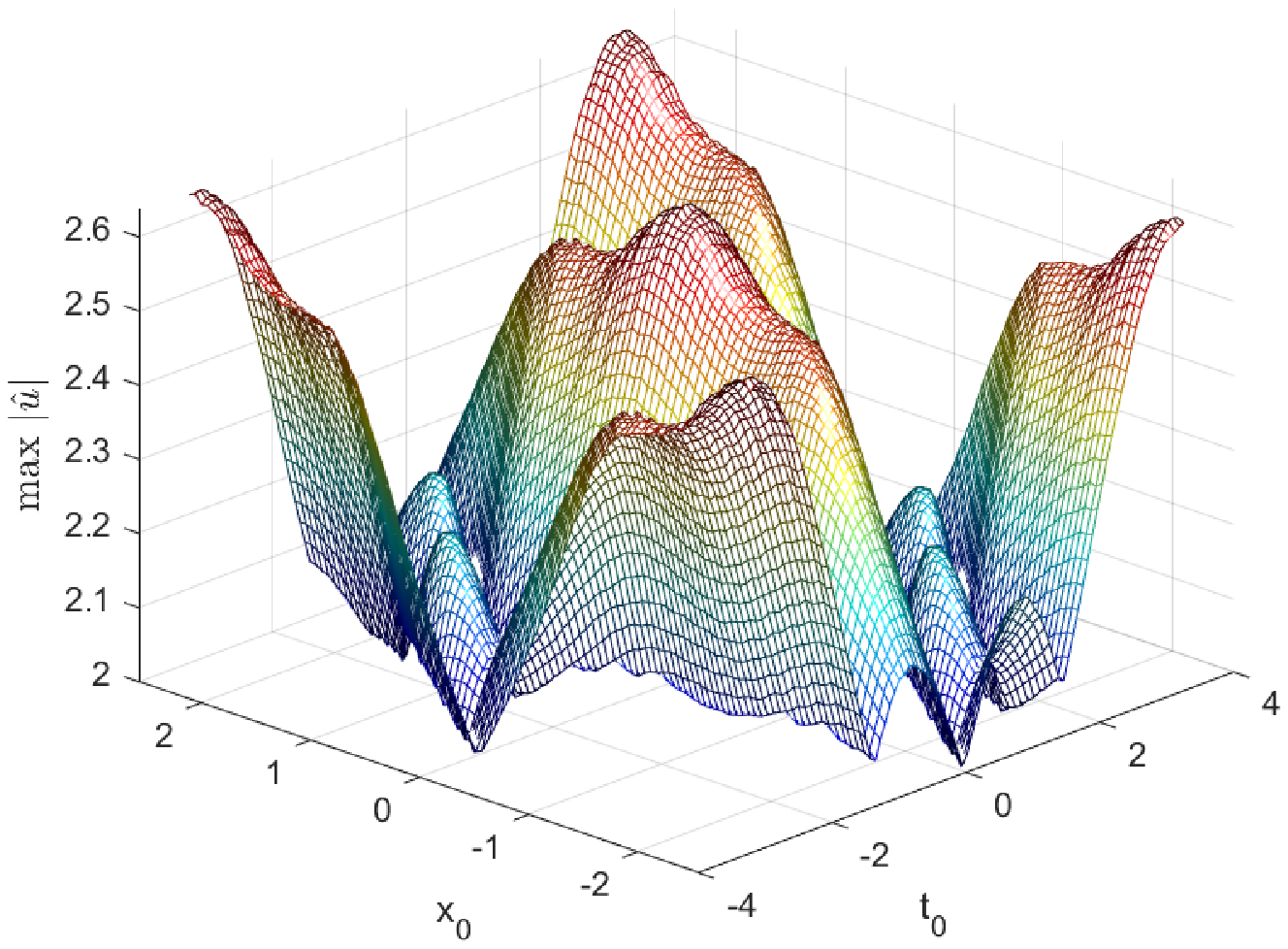}
	\caption{Magnification given by $\max_{(x,t) \in \mathbb{R}^2} |\hat{u}(x,t)|$ versus $(x_0,t_0)$
	for rogue waves generated from the double-periodic solution (\ref{solA}) with
	$k=0.8$ for $\lambda_{1} = \sqrt{ z_{1} }$ (left) and $k = 0.2$ for
    $\lambda_{2} = \sqrt{\xi + i \eta}$ (right).}
	\label{fm4}
\end{figure}

Figure \ref{fm4} shows how $\max_{(x,t) \in \mathbb{R}^2} |\hat{u}(x,t)|$ depends on $(x_{0},t_{0})$
for two rogue waves on Figure \ref{f9} with largest magnification. The maximal
magnification is reached at
$(x_0,t_0) = (0,0)$ or $(x_0,t_0) = (L/2,T/2)$. We have checked numerically
that both points are equivalent to each other. This chess pattern of the maximal magnifications
resembles the chess pattern of maxima in the double-periodic background (\ref{solA}) shown
on Fig. \ref{f1} (right). Note that if $\max_{(x,t) \in \mathbb{R}^2} |\hat{u}(x,t)|$
is computed for a rogue wave with smaller magnification (e.g. for the rogue wave on the top right panel
of Fig. \ref{f9}), then there exist many points of the maximal magnification, which are nearly
equivalent to each other.

We can finally address the question on multiplication factors for the rogue waves on the double-periodic
background. Since $\tilde{u}$ is a translated version of $u$, it follows from (\ref{asymptotic-zero}) that
\begin{equation}
\label{def-10}
|\hat{u}(x_0,t_0)| \leq 3 \max_{(x,t) \in \mathbb{R}^2} |u(x,t)|.
\end{equation}
Let us recall from Figures \ref{fm1} and \ref{fm4} that the maximal magnification of $|\hat{u}|$ is reached
at $(x_0,t_0) = (0,0)$ and
the argument of $\max_{(x,t) \in \mathbb{R}^2} |\hat{u}(x,t)|$ with $(x_0,t_0) = (0,0)$ occurs at $(x,t) = (0,0)$.
If the magnification factor of the rogue wave is defined by
\begin{equation}
\label{def-1}
M_1 := \frac{\max_{(x,t) \in \mathbb{R}^2} |\hat{u}(x,t)|}{\max_{(x,t) \in
\mathbb{R}^2} |u(x,t)|},
\end{equation}
then it follows from (\ref{def-10}) that $M_1 \leq 3$. In fact,
$M_1$ does not exceed the double factor for all rogue waves constructed on Figures \ref{12rogue} and \ref{f9},
as is shown in Table \ref{Table3}.

On the other hand, if the magnification factor is defined by
\begin{equation}
\label{def-2}
M_2 := \frac{\max_{(x,t) \in \mathbb{R}^2} |\hat{u}(x,t)|}{{\rm mean}_{(x,t) \in
\mathbb{R}^2} |u(x,t)|},
\end{equation}
as in physical experiments (see \cite{Tovbis1}), then
$M_2$ exceeds the triple factor for all rogue waves constructed on Figures \ref{12rogue} and \ref{f9}
as shown in Table \ref{Table3}. Thus, all rogue waves constructed here correspond to physically acceptable rogue waves on the double-periodic
background.


\begin{table}[ht]
	\begin{center}
		\begin{tabular}{|c|c|c|c|}
			\hline
			Rogue wave  & Solution & $M_1$ in (\ref{def-1}) & $M_2$ in (\ref{def-2})  \\
			\hline
			$\lambda_1 = \sqrt{z_1}$ & \eqref{solB} with $k = 0.9$ &$1.45$ &
			$3.96$
			\\
			$\lambda_2 = \sqrt{z_2}$ & same &$1.71$ & $4.68$ \\
			$\lambda_3 = \sqrt{z_3}$ & same &$1.84$ & $5.03$ \\
			$\lambda_1 = \sqrt{z_1}$ & \eqref{solA} with $k = 0.8$ &$1.80$ &
			$4.67$ \\
			$\lambda_2 = \sqrt{\xi + i \eta}$ & same &$1.60$ & $4.15$ \\
			$\lambda_1 = \sqrt{z_1}$ & \eqref{solA} with $k = 0.2$ &$1.58$ &
			$3.55$ \\
			$\lambda_2 = \sqrt{\xi + i \eta}$ & same &$1.71$ & $3.84$ \\
			\hline
		\end{tabular}
	\end{center}
	\caption{Magnification factors for the rogue waves constructed on Figures \ref{12rogue} and \ref{f9}.}
	\label{Table3}
\end{table}

~\newpage
\section{Conclusion}
\label{sec-conclusion}

We have constructed analytically and studied numerically the rogue wave patterns appearing on the background
of the double-periodic solutions to the focusing NLS equations. The analytical part relies on the algebraic method
with two eigenvalues and fully characterizes eigenvalues and squared eigenfunctions of the Lax equations
associated with the double-periodic solutions. The numerical part explores explicit representations of
eigenfunctions of the Lax equations and the rogue waves in terms of integrals computed from the double-periodic solutions. We have argued that
the properly defined magnification factor exceeds the triple value for all rogue waves appearing on the
double-periodic background.

This work opens up a number of new directions in the study of rogue waves modeled by the focusing NLS equation.
First, rogue waves on the continuous wave background were observed experimentally either in water tanks
or in laser optics \cite{experiments1}, hence experiments on the double-periodic background are feasible.
Second, rogue waves on the double-periodic background could be relevant to diagnostics
of rogue waves on the ocean surface \cite{CalSch}. Third, it may be interesting to develop the algebraic method
further and to provide the explicit characterization of eigenvalues and squared eigenfunctions for the most general
solution of the third-order and higher-order Lax--Novikov equations. The latter problem is related to characterization
of parameters of the Lax--Novikov equations in terms of parameters of the Riemann Theta function of genus two and higher.

\vspace{0.5cm}

{\bf Acknowledgements.}  Analytical work on this project was supported by the National
Natural Science Foundation of China (No. 11971103). Numerical work was supported by
the Russian Science Foundation  (No.19-12-00253).
	
\setcounter{equation}{0}
\renewcommand{\theequation}{\thesection.\arabic{equation}}

\appendix
\section{Derivation of the explicit solution (\ref{solB}) from (\ref{Jacobi-1-z}) and (\ref{Jacobi-1})}
\label{sec:A}

Setting $z_1+z_2=z_3$ and $z_3=1$ in the solution (\ref{Jacobi-1-z})
yields $\mu = 2 z_2$, $k = z_1/z_2$, and
\begin{equation}
\label{A.1}
z(t) = \frac{z_1{\rm sn}^2(\mu t;k)}{z_2 + z_1{\rm sn}^2(\mu t;k)}.
\end{equation}
By using the ascending Landen transformation (see 16.14 in \cite{AS}),
$$
{\rm sn}(\mu t;k) = \frac{\mu {\rm sn}(t;k_0) {\rm cn}(t;k_0)}{{\rm dn}(t;k_0)},
\quad k_0 := \frac{2 \sqrt{k}}{1+k} = 2 \sqrt{z_1z_2},
$$
the solution (\ref{A.1}) is transformed to the form
\begin{equation}
\label{A.2}
z(t)=\frac{k_0^2{\rm sn}^2(t;k_0){\rm cn}^2(t;k_0)}{1-k_0^2{\rm sn}^4(t;k_0)},
\end{equation}
so that we extract smoothly the square root and obtain
\begin{equation}
\label{A.2-delta}
\delta(t) = \frac{k_0 {\rm sn}(t;k_0){\rm cn}(t;k_0)}{\sqrt{1-k_0^2{\rm sn}^4(t;k_0)}}.
\end{equation}

In order to express $Q_{1,2}$ and $Q_{3,4}$ from $z(t)$ and $\dot{\delta}(t)$ by (\ref{Q-roots-correct}),
we obtain directly
\begin{equation}
\label{A.5}
\sqrt{z_3-z} = \frac{{\rm dn}(t;k_0)}{\sqrt{1-k_0^2{\rm sn}^4(t;k_0)}}
\end{equation}
and
\begin{equation}
\label{A.5a}
\sqrt{z_1 + z_2 - 2 z \pm \frac{\dot{\delta}}{\sqrt{z_3-z}}} = \frac{\sqrt{1 \pm k_0} [1 \mp k_0{\rm sn}^2(t;k_0)]}{\sqrt{1-k_0^2{\rm sn}^4(t;k_0)}},
\end{equation}
where the last equality follows from (\ref{A.2}), (\ref{A.2-delta}), and (\ref{A.5}) which imply
\begin{eqnarray*}
z_1 + z_2 - 2 z \pm \frac{\dot{\delta}}{\sqrt{z_3-z}} & = & 1 - \frac{2 k_0^2 {\rm sn}^2(t;k_0) {\rm cn}^2(t;k_0)}{1-k_0^2 {\rm sn}^4(t;k_0)}
\pm \frac{k_0 ({\rm cn}^2(t;k_0) - {\rm sn}^2(t;k_0) {\rm dn}^2(t;k_0))}{1-k_0^2{\rm sn}^4(t;k_0)} \\
& = & \frac{(1 \pm k_0) [1 \mp k_0 {\rm sn}^2(t;k_0)]^2}{1-k_0^2{\rm sn}^4(t;k_0)}.
\end{eqnarray*}
Substituting (\ref{A.5}) and (\ref{A.5a}) back into (\ref{Q-roots}) yields
\begin{equation}\label{A.6}
Q_{1,2} = \frac{{\rm dn}(t;k_0) \pm \sqrt{1 + k_0} (1 - k_0 {\rm sn}^2(t;k_0))}{\sqrt{1-k_0^2{\rm sn}^4(t;k_0)}}
\end{equation}
and
\begin{equation}\label{A.6a}
Q_{3,4} = \frac{-{\rm dn}(t;k_0) \pm \sqrt{1 - k_0} (1 + k_0 {\rm sn}^2(t;k_0))}{\sqrt{1-k_0^2{\rm sn}^4(t;k_0)}}.
\end{equation}
Computing parameters of the expression (\ref{Jacobi-1}) yields $\nu = \sqrt{z_2}$ and $\kappa^2 = 1 - \frac{z_1}{z_2}$.
We shall now use the descending Landen transformation (see 16.12 in \cite{AS}),
\begin{equation}\label{A.7}
{\rm sn}(\nu x, \kappa) = \frac{(1 + \kappa_0) {\rm sn}(\nu_0 x;\kappa_0)} {1 + \kappa_0 {\rm sn}^2(\nu_0 x;\kappa_0)}, \quad
\nu_0 = \frac{\sqrt{1+k_0}}{2}, \quad \kappa_0 = \sqrt{\frac{1-k_0}{1+k_0}},
\end{equation}
thanks to the following formulas:
$$
\kappa_0 := \frac{1 - \sqrt{1-\kappa^2}}{1 + \sqrt{1 - \kappa^2}} = \frac{\sqrt{z_2} - \sqrt{z_1}}{\sqrt{z_2} + \sqrt{z_1}} =
\frac{1 - \sqrt{k}}{1 + \sqrt{k}} = \sqrt{\frac{1 - k_0}{1 + k_0}}
$$
and
$$
\nu_0 := \frac{\nu}{1 + \kappa_0} = \frac{\sqrt{z_2} \sqrt{1+k_0}}{\sqrt{1+k_0} + \sqrt{1 - k_0}} =  \frac{\sqrt{1+k_0}}{2}
$$
where $k_0 = \frac{2 \sqrt{k}}{1 + k} = 2 \sqrt{z_2(1-z_2)}$ has been used.
Substituting (\ref{A.6}) and (\ref{A.6a}) into (\ref{Jacobi-1}) gives
\begin{equation}\label{A-8}
Q(x,t) = \frac{1}{\sqrt{1-k_0^2 {\rm sn}^4(t;k_0)}} \frac{K_1}{K_2},
\end{equation}
where
\begin{eqnarray*}
K_1 &=& 2 {\rm dn}^2(t;k_0) + {\rm dn}(t;k_0) [\sqrt{1 + k_0} (1 - k_0 {\rm sn}^2(t;k_0)) + \sqrt{1 - k_0} (1 + k_0 {\rm sn}^2(t;k_0))] \\
&& + \sqrt{1 - k_0^2} (1 - k_0^2 {\rm sn}^4(t;k_0)) - (1 + k_0)(1 - k_0 {\rm sn}^2(t;k_0))^2 \\
&& - 2 \sqrt{1 + k_0} (1 - k_0 {\rm sn}^2(t;k_0)) [{\rm dn}(t;k_0) + \sqrt{1 - k_0} (1 + k_0 {\rm sn}^2(t;k_0))] {\rm sn}^2(\nu x;\kappa)
\end{eqnarray*}
and
\begin{eqnarray*}
K_2 &=& 2 {\rm dn}(t;k_0) - \sqrt{1 + k_0} (1 - k_0 {\rm sn}^2(t;k_0)) + \sqrt{1 - k_0} (1 + k_0 {\rm sn}^2(t;k_0)) \\
&& + 2 \sqrt{1 + k_0} (1 - k_0 {\rm sn}^2(t;k_0)) {\rm sn}^2(\nu x;\kappa).
\end{eqnarray*}
Substituting (\ref{A.7}) as ${\rm sn}^2(\nu x;\kappa)$ and
performing computations with Jacobian elliptic functions shows that the expressions for $K_1$ and $K_2$
can be factorized as follows:
which can be further simplified to be
\begin{eqnarray*}
K_1 & = & K k_0 \left(\sqrt{1 + k_0} {\rm sn}^2(t;k_0) {\rm dn}(t;k_0) [1-2 \kappa_0^2 {\rm sn}^2(\nu_0 x;\kappa_0)  +
\kappa_0^2 {\rm sn}^4(\nu_0 x;\kappa_0)] \right. \\
&& \phantom{text} \left. + {\rm cn}^2(t;k_0) [1-2 {\rm sn}^2(\nu_0 x;\kappa_0) + \kappa_0^2 {\rm sn}^4(\nu_0 x;\kappa_0)]\right)
\end{eqnarray*}
and
\begin{eqnarray*}
K_2 & = & K \left( \sqrt{1 + k_0} [1-2 \kappa_0^2 {\rm sn}^2(\nu_0 x;\kappa_0) + \kappa_0^2 {\rm sn}^4(\nu_0 x;\kappa_0)] \right.\\
&& \phantom{text} \left. -{\rm dn}(t;k_0) [1-2 {\rm sn}^2(\nu_0 x;\kappa_0) + \kappa_0^2 {\rm sn}^4(\nu_0 x;\kappa_0)] \right),
\end{eqnarray*}
where
\begin{eqnarray*}
K = \frac{2 ({\rm dn}(t;k_0) + \sqrt{1 - k_0})}{(\sqrt{1 + k_0}-\sqrt{1 - k_0}) (1 + \kappa_0 {\rm sn}^2(\nu_0 x;\kappa_0))^2}.
\end{eqnarray*}
We shall now use the half-argument formula (see 16.18 in \cite{AS}),
$$
{\rm sn}^2(\nu_0 x;\kappa_0) = \frac{1 - {\rm cn}(2 \nu_0 x;\kappa_0)}{1 + {\rm dn}(2 \nu_0 x;\kappa_0)},
$$
which yields
$$
1-2 \kappa_0^2 {\rm sn}^2(\nu_0 x;\kappa_0) + \kappa_0^2 {\rm sn}^4(\nu_0 x;\kappa_0) = \frac{2 {\rm dn}(2 \nu_0 x;\kappa_0)
[ 1 - \kappa_0^2 + {\rm dn}(2 \nu_0 x;\kappa_0) + \kappa_0^2 {\rm cn}(2 \nu_0 x;\kappa_0)]}{[1 + {\rm dn}(2 \nu_0 x;\kappa_0)]^2}
$$
and
$$
1-2 {\rm sn}^2(\nu_0 x;\kappa_0) + \kappa_0^2 {\rm sn}^4(\nu_0 x;\kappa_0) = \frac{2 {\rm cn}(2 \nu_0 x;\kappa_0)
[ 1 - \kappa_0^2 + {\rm dn}(2 \nu_0 x;\kappa_0) + \kappa_0^2 {\rm cn}(2 \nu_0 x;\kappa_0)]}{[1 + {\rm dn}(2 \nu_0 x;\kappa_0)]^2}.
$$
As a result, the expression (\ref{A-8}) with $K_1$ and $K_2$ given above simplifies to the explicit expressions:
\begin{equation}\label{A.9}
Q(x,t)=\frac {k_0}{\sqrt{1-k_0^2 {\rm sn}^4(t;k_0)}}
\frac{\sqrt {1+k_0} {\rm sn}^2(t;k_0) {\rm dn}(t;k_0) {\rm dn}(2\nu_0 x;\kappa_0) + {\rm cn}^2(t;k_0) {\rm cn}(2\nu_0 x;\kappa_0)}
{\sqrt{1+k_0} {\rm dn}(2\nu_0 x;\kappa_0) - {\rm dn}(t;k_0) {\rm cn}(2\nu_0 x;\kappa_0)}.
\end{equation}

Computing $\dot{\theta}$ from $\dot{\theta} = z_1 + z_2 + z_3 - 2 z$ yields
\begin{equation}\label{A.3}
\dot{\theta} = 2 (1 - z(t)) = \frac{2 {\rm dn}^2(t;k_0)}{1 - k_0^2 {\rm sn}^4(t;k_0)}.
\end{equation}
We claim that
\begin{equation}\label{A.4}
\theta(t)=t+\arctan \varphi(t),\qquad \varphi(t) := \frac{{\rm sn}(t;k_0){\rm dn}(t;k_0)}{{\rm cn(t;k_0)}}.
\end{equation}
Indeed, by chain rule, we obtain
$$
\dot{\theta} = 1 + \frac{{\rm dn}^2(t;k_0)  - k_0^2 {\rm sn}^2(t;k_0) {\rm cn}^2(t;k_0)}{{\rm cn}^2(t;k_0) + {\rm sn}^2(t;k_0) {\rm dn}^2(t;k_0))},
$$
which recovers (\ref{A.3}). By using the following elementary formulas
$$
\cos(\arctan \varphi) = \frac{1}{\sqrt{1 + \varphi^2}}, \quad \sin(\arctan \varphi) = \frac{\varphi}{\sqrt{1 + \varphi^2}},
$$
we substitute (\ref{A.4}) into (\ref{double-periodic}) and obtain the explicit expression:
\begin{equation}\label{A.11}
\psi(x,t)=\frac {Q(x,t) - \delta(t) \varphi(t) + i \delta(t) + i Q(x,t) \varphi(t)}
{\sqrt{1 + \varphi^2(t)}} e^{it}.
\end{equation}
Substituting (\ref{A.2}), (\ref{A.9}), and (\ref{A.4}) into (\ref{A.11}) results in the explicit solution (\ref{solB})
with $k_0$ and $\kappa_0$ being written again as $k$ and $\kappa$ in (\ref{solB}) for the sake of notations.
\setcounter{equation}{0}
\section{Derivation of the explicit solution (\ref{solA}) from (\ref{Jacobi-2-z}) and (\ref{Jacobi-2})}
\label{sec:B}

Setting $z_1 = 2\xi$ in the solution (\ref{Jacobi-2-z}) yields $\zeta = 1$. By using the scaling
invariance, we can also normalize the solution by $\xi^2 + \eta^2 = \frac{1}{4}$ which yields $\mu = 2$.
Furthermore, $k$ is related to $\xi$ by $k = 2 \xi$. This simplifies the expression (\ref{Jacobi-2-z}) to the form:
\begin{equation}\label{B.1}
z(t)= k \frac{1-{\rm cn}(2t;k)}{2},
\end{equation}
where $k$ is a free parameter. With the help of the double argument formula
(see 16.18.4 in \cite{AS}), the exact solution (\ref{B.1}) can be rewritten in the form:
\begin{equation}\label{B.2}
z(t)=\frac{k {\rm sn}^2(t;k){\rm dn}^2(t;k)}{1-k^2{\rm sn}^4(t;k)},
\end{equation}
so that we extract smoothly the square root and obtain
\begin{equation}
\label{B.2-delta}
\delta(t) = \frac{\sqrt{k} {\rm sn}(t;k){\rm dn}(t;k)}{\sqrt{1-k^2{\rm sn}^4(t;k)}}.
\end{equation}

In order to express $Q_{1,2}$ and $Q_{3,4}$ from $z(t)$ and $\dot{\delta}(t)$ by (\ref{Q-roots-complex-correct}),
we obtain directly:
$$
z_1-z= \frac{k {\rm cn}^2(t;k)}{1-k^2{\rm sn}^4(t;k)}
$$
and
$$
\xi - z = \frac{k (1 - 2 {\rm sn}^2(t;k) + k^2 {\rm sn}^4(t;k))}{2(1-k^2{\rm sn}^4(t;k))}
$$
We claim the following remarkable formula:
\begin{equation}
\label{B.5}
\sqrt{(\xi - z)^2 + \eta^2} \pm (\xi - z) = \frac{(1 \pm k)(1 \mp k{\rm sn}^2(t;k))}{2(1 \pm k{\rm sn}^2(t;k))}.
\end{equation}
Indeed, we have
\begin{eqnarray*}
(\xi - z)^2 + \eta^2 & = & \frac{k^2 (1 - 2 {\rm sn}^2(t;k) + k^2 {\rm sn}^4(t;k))^2 + (1-k^2) (1-k^2{\rm sn}^4(t;k))^2}{4((1-k^2{\rm sn}^4(t;k))^2}\\
 & = & \frac{(1 - 2k^2 {\rm sn}^2(t;k) + k^2{\rm sn}^4(t;k))^2}{4((1-k^2{\rm sn}^4(t;k))^2} \\
 & = & \left[ \frac{(1 \pm k)(1 \mp k{\rm sn}^2(t;k))}{2(1 \pm k{\rm sn}^2(t;k))} \mp
 \frac{k (1 - 2 {\rm sn}^2(t;k) + k^2 {\rm sn}^4(t;k))}{2(1-k^2{\rm sn}^4(t;k))} \right]^2.
\end{eqnarray*}
which proves (\ref{B.5}). As a result, we obtain from (\ref{Q-roots-complex})
that
$$
Q_{1,2} = \frac{\sqrt{k} {\rm cn}(t;k) \pm \sqrt{k+1} (1 - k {\rm sn}^2(t;k))}{\sqrt{1 - k^2 {\rm sn}^4(t;k))}}
$$
and
$$
\alpha = - \frac{\sqrt{k} {\rm cn}(t;k)}{\sqrt{1-k^2{\rm sn}^4(t;k)}}, \quad
\beta = \frac{\sqrt{1 - k}(1 + k{\rm sn}^2(t;k))}{\sqrt{1 - k^2{\rm sn}^4(t;k)}}.
$$
Computing parameters of the expression (\ref{Jacobi-2}) yields
$\nu = \sqrt{2}$ and $2 \kappa^2 = 1-k$. In order to compute $\gamma$,
we obtain
\begin{eqnarray*}
\gamma^2 & = & \frac{(Q_2 - \alpha)^2 + \beta^2}{(Q_1 - \alpha)^2 + \beta^2} \\
& = & \frac{1 - 2k^2 {\rm sn}^2(t;k) + k^2{\rm sn}^4(t;k) + 2k {\rm cn}^2(t;k) - 2 \sqrt{k(1+k)}{\rm cn}(t;k) (1 - k {\rm sn}^2(t;k))}{
1 - 2k^2 {\rm sn}^2(t;k) + k^2{\rm sn}^4(t;k) + 2k {\rm cn}^2(t;k) + 2 \sqrt{k(1+k)}{\rm cn}(t;k) (1 - k {\rm sn}^2(t;k))}
\end{eqnarray*}
which simplifies to
\begin{eqnarray*}
\gamma =  \frac{1 - 2k^2 {\rm sn}^2(t;k) + k^2{\rm sn}^4(t;k) + 2k {\rm cn}^2(t;k) - 2 \sqrt{k(1+k)}{\rm cn}(t;k) (1 - k {\rm sn}^2(t;k))}{1-k^2 {\rm sn}^4(t;k)}
\end{eqnarray*}
and leads to
\begin{equation}\label{B.9}
\gamma + 1 = \frac{2 (1 + k - \sqrt{k(1+k)} {\rm cn}(t;k))}{1 + k {\rm sn}^2(t;k)},\quad
\gamma - 1 = \frac{2 (k {\rm cn}^2(t;k) - \sqrt{k(1+k)} {\rm cn}(t;k))} {1 + k {\rm sn}^2(t;k)}.
\end{equation}
Substituting $Q_{1,2}$ and (\ref{B.9}) into (\ref{Jacobi-2}) yields after some lengthy but direct computations:
\begin{equation}\label{B.11}
Q(x,t) = \frac {k \sqrt{k(1+k)} {\rm cn}(t;k) {\rm sn}^2(t;k) + {\rm dn}^2(t;k) {\rm cn}(\sqrt{2}x;\kappa)}
{\sqrt{1 - k^2{\rm sn}^4(t;k)}\left[\sqrt{1+k} - \sqrt{k} {\rm cn}(t;k) {\rm cn}(\sqrt{2} x;\kappa)\right]}.
\end{equation}
Recall the representation (\ref{double-periodic}) with $Q(x,t)$ in (\ref{B.11}) and
$\delta(t)$ in (\ref{B.2-delta}). Computing $\dot{\theta}$ from $\dot{\theta} = z_1 + z_2 + z_3 - 2 z$ yields
\begin{equation}\label{B.4a}
\dot{\theta} = 2(k-z(t)) = \frac{2 k {\rm cn}^2(t;k)}{1 - k^2 {\rm sn}^4(t;k)}.
\end{equation}
We claim that
\begin{equation}\label{B.4}
\theta(t) = k t + \arctan \varphi(t),\qquad \varphi(t) := \frac{k{\rm sn}(t;k){\rm cn}(t;k)}{{\rm dn}(t;k)}.
\end{equation}
Indeed, by chain rule we obtain
\begin{eqnarray*}
\dot{\theta} & = &  k + \frac{k ({\rm dn}^2(t;k) ({\rm cn}^2(t;k) - {\rm sn}^2(t;k)) +
k^2 {\rm cn}^2(t;k) {\rm sn}^2(t;k))}{{\rm dn}^2(t;k) + k^2 {\rm cn}^2(t;k) {\rm sn}^2(t;k))},
\end{eqnarray*}
which recovers (\ref{B.4a}). Substituting (\ref{B.2}), (\ref{B.11}), and (\ref{B.4})
into
\begin{equation*}
\psi(x,t)=\frac {Q(x,t) - \delta(t) \varphi(t) + i \delta(t) + i Q(x,t) \varphi(t)}
{\sqrt{1 + \varphi^2(t)}} e^{i k t}.
\end{equation*}
yields the explicit solution (\ref{solA}).

\end{document}